\documentclass[twocolumn,showpacs,preprintnumbers,amsmath,amssymb]{revtex4}

\usepackage{graphicx}
\usepackage{dcolumn}
\usepackage{bm}

\begin{document}

\preprint{}

\title{Transition from isotropic to anisotropic beam profiles in a linear focusing channel.}

\author{Wilson Simeoni Jr.}
\email{wsjr@if.ufrgs.br}
\affiliation{Instituto de F\'{\i}sica,
Universidade Federal do Rio Grande do Sul
Caixa Postal 15051, 91501-970, Porto Alegre, RS, Brazil
}

\date{\today}

\begin{abstract}
This paper examines the transition from isotropic to anisotropic beam profiles in a linear focusing channel. Considering a high-intensity ion beam in space-charge dominated regime and large mismatched {\it RMS} beam size initially, observe a fast anisotropy situation of the beam, characterized for a transition of the transversal section round to elliptical with a coupling of transversal emittance driven for instabilities of nonlinear space-charge forces. Space-charge interactions in high-intensity linear accelerator can lead to equipartitioning of energy between the degrees of freedom. The anisotropization phenomena suggest a kind of route to equipartition. In order to understand the initial dynamical behavior of an anisotropic beam, in particular, to study possible mechanisms of equipartition connected with phase space we have to know how we can compute the variables (volume, area of surface, and area projected) that characterize the anisotropic beam in phase space. The beam began in a nonequilibrium state evolved toward a metaequilibrium in which the particle orbits filled an invariant measure of phase space. We shall call this subspace of the phase space \textquotedblleft~$\xi$~\textquotedblright, where $\xi$ is the ratio of oscillations energies in the $x$ and $y$ directions. It is well-know~\cite{Ohnuma1} that an isolated difference ressonance of the form $l\omega_x-m\omega_y=0$, where $\omega_x$ is single particle frequency along of $x$-direction and $\omega_y$ is single particle frequency along of $y$-direction, lead to the motion bounded in both directions and \textquotedblleft~$\xi$~\textquotedblright remains unchanged. The purpose of this paper is to propose one definiton of the anisotropic equipartition~\cite{Yankov1}. Anisotropic equipartition corresponds to a phase space density uniform on the surface invariant of the $\xi$, a version of the ergodic hypothesis where the $\xi$ invariant play the role of the conserved energy~\cite{Kandrup3}. In the state of anisotropic equipartition, the beam temperature is stationary, the entropy grows in the cascade form, there is a coupling of transversal emittance, the beam develops an elliptical shape with a increase in its size along one direction and there is halo formation along one direction preferential.
\end{abstract}

\pacs{41.75.-i, 41.85.Ja, 29.27.Bd, 52.59.Sa, 52.25.Kn, 45.50.Dd}
\keywords{Suggested keywords}
\maketitle

\section{\label{Intro}Introduction}

In space-charge dominated beams the nonlinear space-charge forces produce a filamentation pattern, which results in a 2-component beam consisting of an inner core and an outer halo~\cite{Lagniel1,Lagniel2,Gluckstern1,Jameson1}. When this core is mismatched in a uniform linear focusing channel, the envelope oscillates and the particles, represented by single test particles, oscillate about and through the core. This mechanism is called particle-core model, considering a test particles initially located outside the core. The test particles execute betatron oscillations under the influence of space-charge field induced by the oscillating core, and exhibit various nonlinear behaviors including parametric resonance. In a number of numerical analyses and macro-particle simulations, it has been revealed that the 2:1 parametric resonance between test particle oscillation and breathing core oscillation is a main cause of halo formation~\cite{Wangler1,Gluckstern2,Gluckstern3,Qian1,Okamoto1,Ikegami2,Fedotov1}. 

Space-charge induced coupling between different degrees of freedom can be responsible for emittance growth or transfer of emittance from one phase plane to another. The underlying instability mechanism is coherent if it depends primarily on the electric field due to collective motion~\cite{Hofmann1,Hofmann3}, in contrast with an incoherent that can be described as single-particle effect. In an analytical single-particle analyses Montague~\cite{Montague1} pointed out that the space-charge driven fourth order difference resonance may lead to emittance coupling. The theory of space-charge coupling resonances has been studied most thoroughly by Hofmann using Vlasov equation~\cite{Hofmann2}. Intrinsic coupling resonances are an important consideration in linacs and may also be of interest in rings when the tune separation is small~\cite{Fedotov2}. These resonances are accompanied by halo formation in the receiving energy. In the presence of internal energy anisotropic between different degrees of freedom, initially small space-charge coupling terms can grow exponentially due to collective instability. The coupling resonaces driven by the beam space-charge fields depending only on the relative emittances and average focusing strengths. In anisotropic beams, the emittance and/or external focusing force strength are different in the two transverse directions. Ikegami~\cite{Ikegami1} deal with effects of anisotropic of beam cores on halo dynamics.

Most of the halo studies so far have considered round beams with axisymmetric focusing. Some news aspects caused by anisotropy demonstrate an influence of the mismatch on halo size~\cite{Hofmann1,Kishek1}. The mismatch oscillations can drive particles into a halo as a result of resonant interaction of these particles with the mismatch mode~\cite{Gluckstern2}. So far only second order round beams have been considered as possible mismatch modes: the influence of anisotropic on second and higher order mismatch modes is expected to be an important factor in halo formation.

Previous work has been done in a nonlinear analyses of the beam transport considering nonaxisymmetric perturbations~\cite{Simeoni1}. In which it is shown that large-amplitude breathing oscillating of an initially round beam couple nonlinearly to quadrupole-like oscillations, such that the excess energy initially constrained to the axisymmetric breathing oscillating is allowed to flow back and forth between breathing and quadrupole-like oscillations. In this case, the beam develops an elliptical shape with a increase in its size along one direction as the beam is transported. This is a higly nonlinear phenomenon that occurs for large mismatch amplitudes on the order of 100\% ~\cite{Lapostolle1}.

This papers~\cite{Simeoni2} deals with the coupled motion between the two tranverse coordinates of a particles beam arising from the space-charge forces, and, in particular, with the effects of the beam coupling which have ratio $r_x/r_y = 1$, $r_x$ and $r_y$ are the envelope semi-axes-rms. A beam with nonuniform charge distribution always gives rise to coupled motion. However, it is only when the ratio $r_x/r_y = 1$ and the large beam size-rms mismatched on the order of 100\% ~\cite{Lapostolle1} that the coupling can produce an observable effect in the beam as a whole. This effect arises from both, a beating in amplitude between the two coordinate directions for the single-particle motion, and from the coupling between oscillations modes beam, resulting in growth and transfer of emittance from one phase plane to another, in the beam develops an elliptical shape and therefore in the transition from isotropic to anisotropic beam profiles. 

Nonlinear space-charge forces can also lead to equipartitioning of energy between degrees of freedom. The question is how a system of collisionless particles coupled by long-range space-charge forces will equipartition.~\cite{Lagniel3}. In the presence of nonlinear coupling mechanisms, the beam can be predicted to equipatition, or reach a state where velocity spreads in two directions are equal. In space-charge dominated beams, Coulomb collisions are infrequent to account for the energy transfer, whereas space charge waves have been shown to be a possible coupling canditate~\cite{Wangler2,Kishek1,Kishek2}. Equipartitioning of anisotropic beam involves nonlinear energy transfer and evolution towards a metaequilibrium state, as a consequence of resonant phase mixing~\cite{Bohn1,Kandrup1,Kandrup2}. Strictly speaking, resonant phase mixing is a reversible process in which it is governed by Vlasov's equation. However, an essential question for the accelerator designer is wheter this process in operationally reversible. While it may be possible in principle to compensate operationally against phase space dilution, this compensation must be completed before any mixing has smeared a significant number of particles through global regions of phase space. It arises regarding any process for manipulating a beam with space-charge~\cite{Yu1}, be this changing the beam's tranverse geometry (round-to-elliptical or elliptical-to-round transformations~\cite{Simeoni1,Brinkmann1}). We are working toward quantifying the relationship between the anisotropy of the beam and the equipartition. The equipartition of beam is driven for anisotropics processes. It is, therefore, to develop an improved understanding of fundamental collective stability properties, including the case where a large temperature anisotropy can drive electrostatic Harrys-type~\cite{Harris1} and/or eletromagnetic Weibel-type~ \cite{Weibel1} instabilities, familiar in the study of electrically neutral plasmas~\cite{Startsev1}. In plasmas with strongly anisotropic distribution functions, collective instabilities may develop if there is sufficient coupling between the degrees of fredom. Previous studies have mostly focused on the electrostatic Harris-type anisotropic-driven instability for beams~\cite{Startsev2}. It has been shown that a fast, electrostatic instability develops, and satures nonlinearly, for sufficiently temperature anisotropic.

The term equipartition broadly refers to the ergodic property of multi-dimensional Hamiltonian systems, which tend to distribute uniformly over the phase space surface of constant energy. The conservation of energy plays the fundamental role in classical equilibrium thermodynamics. The term \textquotedblleft turbulent equipartiton \textquotedblright was introduced by Yankov~\cite{Yankov2} in order to describe the turbulent relaxed state, in which the system assumes a uniform distribution on the surface of constant invariants respected by turbulence. In plasmas physics, the best known example of a turbulent equipartition is the quasilinear plateau of the distribution function caused by the nonlinear Landau damping of plasmas waves. In a toroidal turbulent plasma, the relevant invariants are given by the frozen-in law (in the fluid limit) or the adiabatic invariant and the Liouville theorem (in the collisionless limit), both limits derivable from the more general Poincare invariant. The plasma mixing by low-frequency electrostatic modes in a Tokamak, subject to these conservative laws, results in the inhomogeneous density and temperature profiles peaked at the center even in the absence of particles and energy fluxes, thus presenting the underlying mechanism of the pinch effect and the profiles consistency in Tokamak~\cite{Yankov1,Isichenko1}. The purpose of this paper is to propose one definiton of the anisotropic equipartition. Anisotropic equipartition corresponds to a phase space density uniform on the surface invariant of the ($\xi=(r_y\epsilon_x)^2/(r_x\epsilon_y)^2$), where $\xi$ is the ratio of energy oscillations in the $x$ and $y$ directions, a version of the ergodic hypothesis where the  $\xi$ invariant play the role of the conserved energy~\cite{Kandrup3}. In the state of anisotropic equipartition, the temperature is stationary, the entropy grows in the cascade form, there is a coupling of transversal emittance, the beam develops an elliptical shape with a increasing in its size along one direction and there is halo formation along one direction preferential.

This paper is organized as follow. In Sec.~\ref{EQM} the models equations are derived. In Sec.~\ref{IsoAniso}  we examine the transition from isotropic to anisotropic beam profiles in a linear focusing channel and the transition of the transversal section from round to elliptical, with a coupling of transversal beam emittance. In Sec.~\ref{Equiparti} we study possible mechanisms of equipartition connected with phase space and compute the variables (volume, area of surface, and area projected) that characterize the anisotropic beam in phase space. Finally in Sec.~\ref{Conclu} we discuss implications and possibles extensions of our results.

\section{\label{EQM}The model equations}
We consider an axially long unbunched beam of ions of charge $q$ and mass $m$ propagating with average axial velocity $\beta_bc{\bf \hat e}_z$ ($c$ is the speed of light {\it in vacuo} and relativistic factor $\gamma_b=1/\sqrt{1-\beta^{2}_{b}}$) along an uniform linear focusing channel, self-field interactions are electrostatic. The beam is assumed to have an elliptical cross section centered at $x=0=y$ and vanishing canonical angular momentum $P_{\theta}\equiv \langle xy^{'}-yx^{'}\rangle =0$, where $x$ and $y$ are the positions of the beam particles. We consider nonuiform density beam in space-charge dominated regime and $\epsilon_x=\epsilon_y$ emittance, initially. The general property of space-charge dominated beam behaviour is that a beam with an initial nonlinear profile tends to be more uniform and this process is associated with strong emittance growth and the appearance of beam halo.

As demonstrated by Sacherer~\cite{Sacherer1} and Lapostolle~\cite{Lapostolle1} enevelope equations for a continuous beam are not restricted to uniformly charged beams, but are equally valid for any charge distribution with elliptical symmetry, provided the beam boundary and emittance are defined by {\it rms} (root-mean-square) values. Thus, we consider the parabolic density beam ($n_b=2N_b/\pi r_xr_y\left[1-x^2/r^2_x-y^2/r^2_x\right]$) where $N_b$ is the axial line density, $r_x=\sqrt{6\langle x^2\rangle}$ and $r_y=\sqrt{6 \langle y^2\rangle}$  are ellipsis semi-axes {\it rms}. A main point is that for parabolic density distribution the fourth order space-charge potential driving the coupling is already present in the initial distribution, hence emittance exchange and beam develops an elliptical shape immediately.

For a parabolic density $n(x,y)$, Poisson's equation $\nabla^{2}_{\bot}\phi=-qn/\varepsilon_0$, $\varepsilon_0$ is the permittivity of free space, provides the basis for obtaining the space-charge field component (assuming paraxial approximation). The density is assumed to be zero outside of the ellipse.The solution has been given by Lapostolle~\cite{Lapostolle2}. The electrostatic potential $\phi$ is given by :

\begin{widetext}
\begin{equation}
\label{eq1}
\phi_{in}=\frac{2qN_b}{\pi\varepsilon_o}\Biggl\{\frac{x^2}{r_x(r_x+r_y)}+\frac{y^2}{r_y(r_x+r_y)} 
-x^4\Biggl[\frac{2r_x+r_y}{3r_x^3(r_x+r_y)^2}\Biggr]-y^4\Biggl[\frac{2r_y+r_x}{3r_y^3(r_x+r_y)^2}\Biggr] - x^2y^2\Biggl[\frac{1}{r_xr_y(r_x+r_y)^2}\Biggr]\Biggr\}
\end{equation}
\end{widetext}
inside the beam and

\begin{widetext}
\begin{equation}
\label{eq2}
\phi_{out}=\frac{q}{4\pi\varepsilon_0}\log\Biggl[y^2+x^2 +\lambda+\sqrt{2}y\Delta^{+}+\sqrt{2}x\Delta^{-}\Biggr]
+\frac{q}{2\pi\varepsilon_0\lambda^2}\Biggl[y^2-x^2-\frac{y}{\sqrt{2}}\Delta^{+}+\frac{x}{\sqrt{2}}\Delta^{-}\Biggr] 
\end{equation}
\end{widetext}
outside the beam, where $\lambda=r_x^2-r_y^2,$~$\Lambda=\sqrt{(x^2-y^2-\lambda^2)^2+4x^2y^2}$~and $\Delta^{\pm}=\sqrt{\Lambda\pm(x^2-y^2 \mp\lambda^2)}$.

The transverse orbit $x(s)$ of a beam particle satisfy the paraxial equation of motion 
\begin{equation}\label{eq3}
x^{''}+\kappa_0^2x=\frac{-q}{m\gamma_b\beta_b^2c^2}\frac{\partial \phi}{\partial x}
\end{equation}
with an analogous equation for orbit $y(s)$. Here, $s$ is the axial coordinate of a beam, primes denote derivates with respect to $s$ and $\kappa_0$ is represented constant focusing force.

The envelope of the beam is an elliptical cross-section with {\it rms} radii $r_j$ (henceforth, $j$ ranges over both $x$ and $y$) that obey the {\it rms-KV} envelope equations~\cite{Davidson1}
\begin{equation}\label{eq4}
r^{''}_{j}+\kappa_0^2r_j-\frac{2K}{r_x+r_y}-\frac{\epsilon^2_j}{r^3_j}=0
\end{equation}
Here, $K=q^2N_b/\pi^2\varepsilon_0\gamma_b^3\beta_b^2mc^2$ is the dimensionless perveance of the beam. $\epsilon_j$ is {\it rms}-emittance of the beam along the $j$-plane. 

The $\epsilon_j=\sqrt{\langle j^{2}\rangle \langle j^{'2}\rangle -\langle jj^{'}\rangle ^2}$ can analytical been calculated following a model proposed to Lapostolle {\it et al.}~\cite{Lapostolle3} for nonlinear space-charge forces, and a nonlinear analysis proposed to Pakter {\it et al.}~\cite{Simeoni1} for large mismatch beam. According to Lapostolle {\it et al.}~\cite{Lapostolle3} nonlinear space-charge forces cause a change in the momentum components, which is equal to the product of the force and the time over which the force acts. The force depends on the spatial distribution of the particles, and particle coordinates on which the force acts. In general, these changes in the momentum components change the phase-space distribution of the particles. Thus the particles experience a space-charge impulse, but do not propagate far enough for their positions to change appreciably. For a beam with large mismatch amplitudes, Pakter {\it et al.}~\cite{Simeoni1} has demonstrated that mismatch is like an impulse so fast that particle's positions not change, only momentum suffers a discontinuous variation. Mismatched oscillatories modes beam oscillate between a maximum value and a minimum value around matched envelope with a given periodicity. To label different mismatched oscillations, its define a mismatch amplitude as $A\equiv\psi_{max}/\psi_{0}\geq 1$ where $\psi_{max}$ is maximum oscillation amplitude of breathing and quadrupole modes, and  $\psi_{0}$ is matched oscillation amplitude. Mismatched oscillations are excess energy given to the beam. In particular, there is a threshold mismatched amplitude $A_{th}$ above which effective energy exchange between breathing and quadrupole modes takes place~\cite{Simeoni1}.

From electrostatic potential and mismatch amplitudes on the order of $100\%$, the tranverse momentum impulse can be calculated. This results in a new phase-space distribution and new {\it rms} emittance~\cite{Neri1, Dragt1}. For example, in the $x$ plane; the change in the momentum component is $\triangle p_x=qE_xA/v_b$, where $E_x=\partial \phi^{in}_{x}/\partial x$ is electric field, $v_b$ is the beam velocity. The impulse can also be expressed as change in the divergence angle, given non-relativistically in the paraxial aproximation by $\triangle x^{'}=qE_xA/m_bv^2_b$. If the second moments of the particle distribution can be evaluted from the expression for $x$ and $x^{'}$, the {\it rms} emittance can be obtained. Knowing that divergence is $x^{'}=x/r_x+qE_xA/m_bv^2_b$, the {\it rms} emittance for parabolic density yields~\cite{Simeoni2}
\begin{equation}
\label{eq5}
\epsilon_x=\frac{1}{90}\sqrt{15}KA\left \{\frac{\left(\frac{r_x}{r_y}\right)^2\left[5\left(\frac{r_x}{r_y}\right)^2+2\frac{r_x}{r_y}+5\right] }{\left(1+\frac{r_x}{r_y}\right)^4}\right\}^{1/2}
\end{equation}
the result is easily transformed to the $y$ plane interchanging $r_x$ and $r_y$. It observes that emittance depends of perveance beam $K$, mismatched amplitude of oscillatories modes beam $A$ and ratio of the beam semi-axes. The first term corresponds to the filamention effect caused by the fourth order term in the electrostatic potential. The second term comes from the coupling, i.e the dependence of the $x$ component of the potential on the $y$ coordinate, which produces the spreading of the initial filamention. The last term is a cross term between the filamention term and the coupling term.

KV distribution has frequently been taken as a theoretical basis of particle-core model because space-charge forces are linear. However, there is no doubt that realistic intense beams contain a fully nonlinear nature. It may thus be reasonable to try constructing an alternate particle-core algorithm with a nonlinear core potential even if the model is only aproximation. From this point of view, we introduce here a parabolic core under a simplifying assumption . In this particle-core model, the core is described by {\it rms} envelope equation~(\ref{eq4}), and the halo particles are modeled using test particles that subject to the external force and the time-dependent nonlinear space-charge force associated with the parabolic core. We assume that the parabolic-type density is roughly maintained even for a mismatched beam. Thus the spatial distribution is assumed to remain unchanged as the beam propagate. The test particles do not affect the motion of the core~\cite{Piovella1} and they are described by equation~(\ref{eq3}) taking $\phi_{in}$ for inside beam, and $\phi_{out}$ for outside beam. Thus the theory is not self-consistent.

\subsection{\label{IsoAniso}Transition from isotropic to anisotropic beam}
It is easy to verify that there is a particular solution of the envelope equations (\ref{eq4}) for which $r_j(s)=r_{b0}= \left[\left (K+(K^2+4\kappa_0^2\eta^2)^{1/2}\right)/ 2\kappa_0^2\right]^{1/2}$, where $\eta=\epsilon_x/\epsilon_y$ is ratio emittance. This corresponds to the so called matched solution for which a circular beam of radius $r_{b0}$ preserves its shape throughout the transport along the focusing channel. Then, we transform the equations to a dimensionless form introducing the following dimensionless variables an parameters: $\tau=\kappa_0s$ for the independent variable, $\tilde{r_x}=\sqrt{\kappa_0/\epsilon_y}r_x$ and $\tilde{r_y}=\sqrt{\kappa_0/\epsilon_y}r_y$ for envelope beam, $\tilde{x}=\sqrt{\kappa_0/\epsilon_y}x$ and $\tilde{y}=\sqrt{\kappa_0/\epsilon_y}y$ for test-particle, and $\tilde{K}=K/\epsilon_y\kappa_0$ for the scaled space-charge perveance. In addition, we introduce the following anisotropy variables: the ratio emittance $\eta$, ratio of the envelope beam $\chi=r_x/r_y$ and the mismatch factor $\nu=r_x/r_{b0}=r_y/r_{b0}$.

We launch the beam with $\tilde{K}=3$, $\kappa_0=1$, $\nu=2.4$,~$A=2.0$,~$r_{j0}=\nu r_{b0}$,~$r_{x0}=\,r_{y0}$ and $\eta=1$ initially, and integrate firstly the envelope equations~(\ref{eq4}) up to $s=50.0$. The corresponding evolution of the $rms$-emittance~(\ref{eq5}) is used for $\epsilon_j$ in Eq.(\ref{eq4}). It is convenient to introduce new canonical variables defined as $X_s=(r_x+r_y)/2$ and $X_a=(r_x-r_y)/2$ to analyze oscillations modes beam. Note that $X_s$ are oscillations where $r_x(s)$ and $r_y(s)$ oscillate in phase : breathing modes. And $X_a$ are oscillations where $r_x(s)$ and $r_y(s)$ oscillate with opposite phase : quadrupole modes~\cite{Lund1}.

\begin{figure}[htb]
\centering
\includegraphics*[width=85mm]{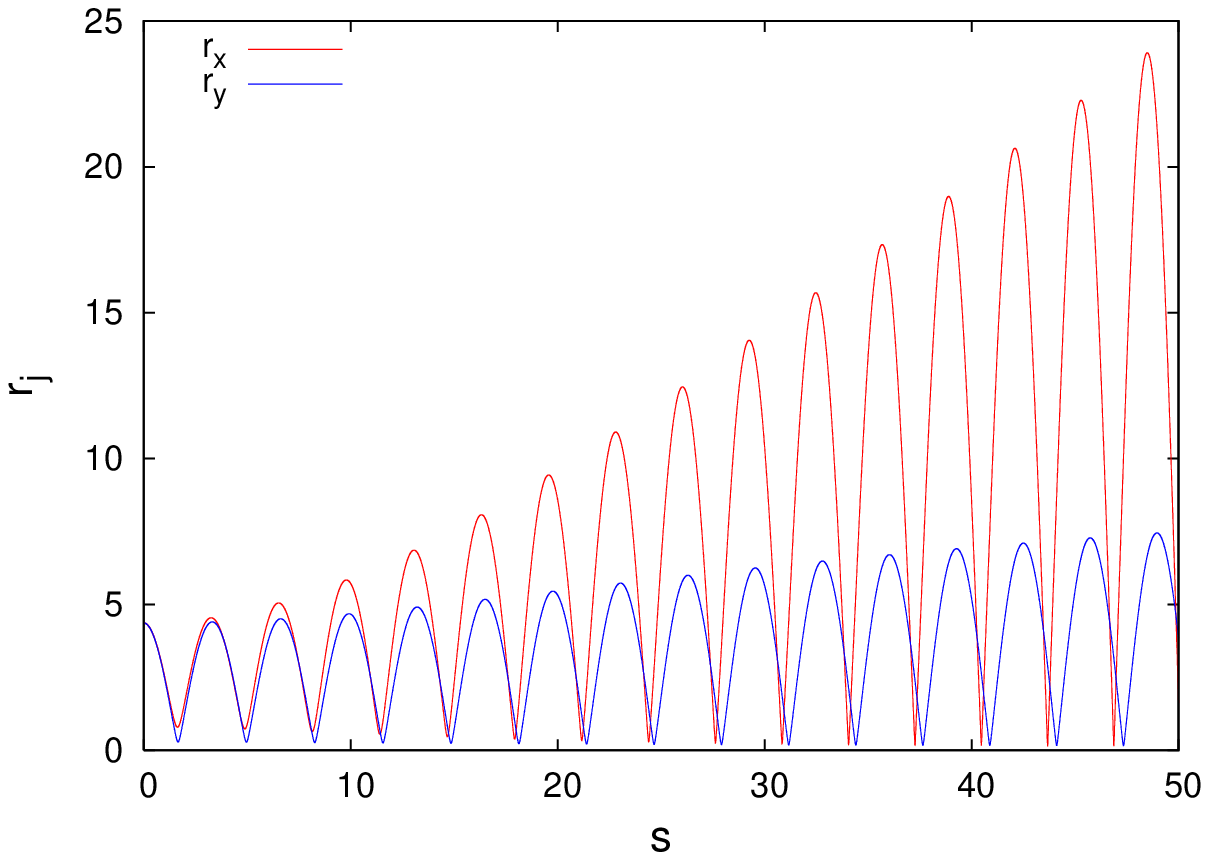} \\
\includegraphics*[width=85mm]{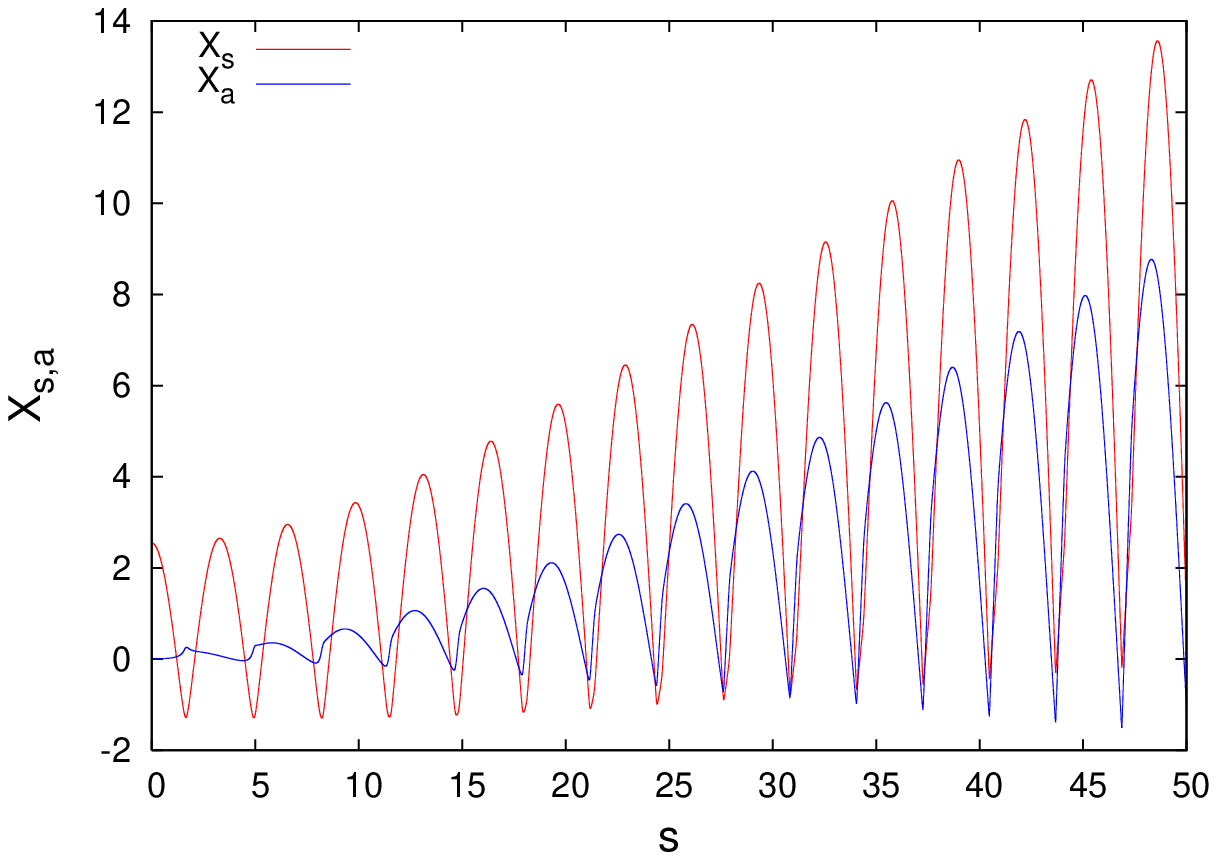}
\caption{Evolution of the envelopes obtained by direct integration of the envelope equations~(\ref{eq4}) (top). $r_x$ and $r_y$ are represented by red and blue lines, respectively. Evolution of the oscillations modes beam (bottom). $X_s$ and $X_a$ are represented by red and blue lines, respectively.}
\label{core}
\end{figure}

In Fig.~\ref{core} the beam develops an elliptical shape with a increase in its size along of $x$-direction~\cite{Simeoni2}. This effect is caused for large beam size-rms mismatched its that coupled the oscillations modes beam, perturbing nonlinear space-charge force~\cite{Strasburg1}. Therefore perturbation induction coupling between different degrees of freedom and core-core resonances~\cite{Hofmann1, Wangler2}. Real core-core resonances~\cite{Hofmann2} are observed for {\it rms} mismatched beam fulfilling internal resonance conditions between the planes. Beam size-rms mismatched is excess energy given to the system. In general, this excess energy appear as oscillation energies in the degrees of freedom of the system. Space-charge couple some degrees of freedom causing resonance between them. Coupling resonance leading to an exchange of energy between the relevant degrees of freedom. In Fig.~\ref{core} (bottom) there is an increase of oscillation amplitude of modes featuring a resonance between breathing and quadrupole modes. To analyse this resonance we computed dimensionless frequencies of breathing and quadrupole modes using Fourier analyses these modes. The frequency associated with the breathing or quadrupole mode corresponds to the maximum in the Fourier transform. We used FFTW (The Fastest Fourier Transform in the West) to compute numerically the frequencies. Features this method are found on the web page~\cite{web} and papers~\cite{FFTW}. Breathing mode frequency is $\omega_{X_s}=2.01334$ and quadrupole mode frequency is $\omega_{X_a}=2.01334$. Note that both modes has the same frequency. Therefore breathing mode and quadrupole mode are resonance. This core-core resonance together with single-particle resonances are making the beam to develop an elliptical shape with a increase in its size along of $x$-direction~\cite{Hofmann2}. It should be noted that in the anisotropic case ($\chi\neq1$ and $\eta\neq1$) both the breathing and quadrupole modes have quadrupolar symmetry. 

\begin{figure}[htb]
\centering
\includegraphics*[width=85mm]{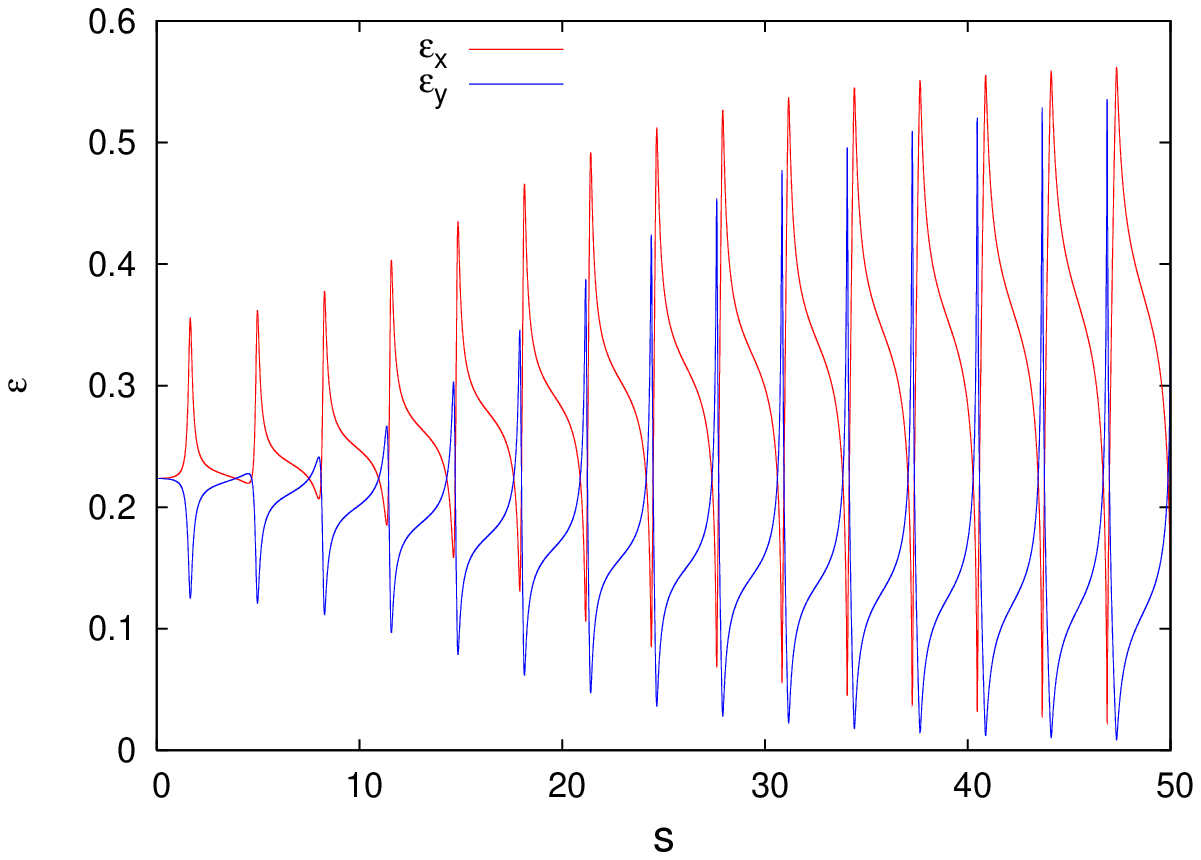}\\
\includegraphics*[width=85mm]{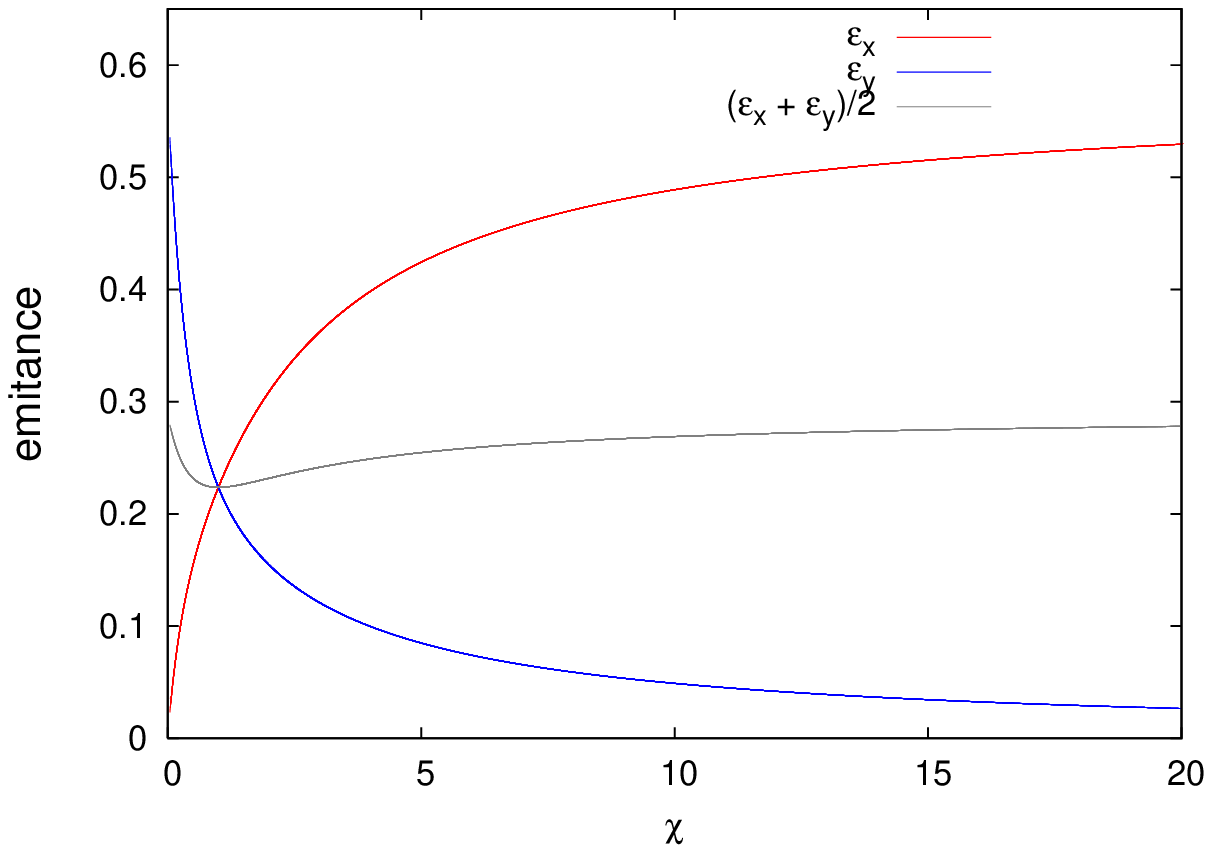}
\caption{Evolution of the emittance (top) obtained of the equation~(\ref{eq5}). $\epsilon_x$ and $\epsilon_y$ are represented by red and blue lines, respectively. Emittance and sum transversal emittance (bottom) as a function of the ratio of the envelope beam $\chi=r_x/r_y$. $\epsilon_x$,~$\epsilon_y$ and $(\epsilon_x+\epsilon_y)/2$ versus $\chi$ are represented by red, blue and gray lines, respectively. }
\label{emit}
\end{figure}

\begin{figure}[htb]
\centering
\includegraphics*[width=42mm]{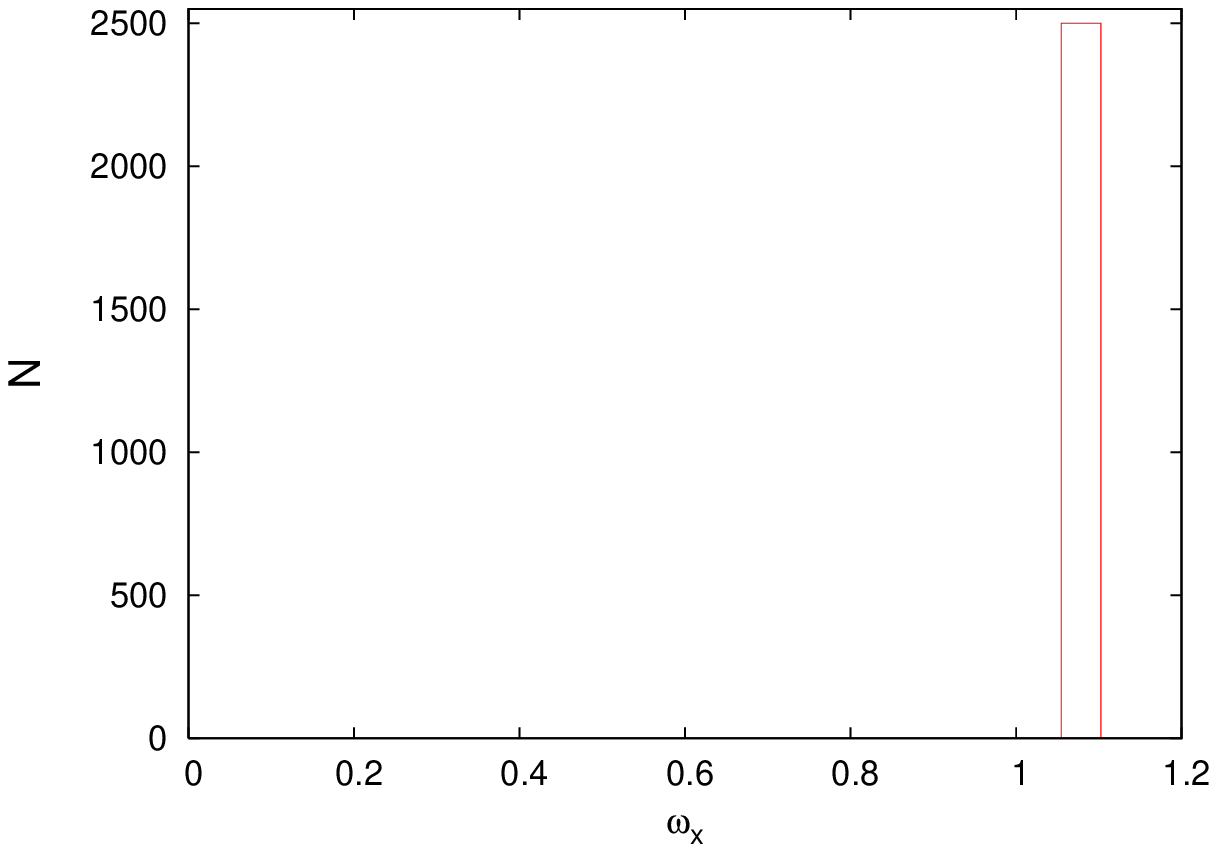} 
\includegraphics*[width=42mm]{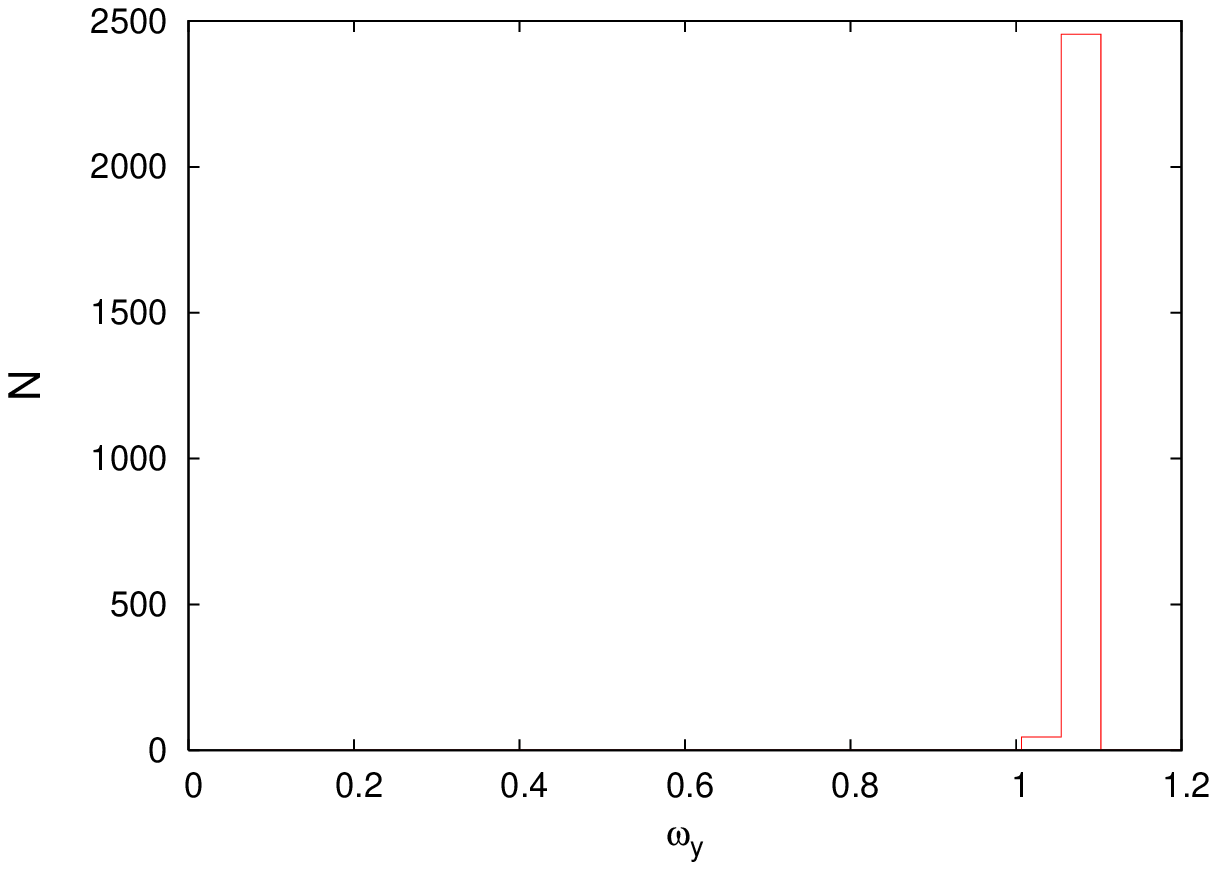}
\caption{Histogram  of the dimensionless frequencies $\omega_x$ and $\omega_y$ of $2500$ test particles. We launch the test particles along the $x$- and $y$-axes of beam in specified region inside the beam (between $0.01r_x$ and $0.7r_x$ spaced by $0.0004r_x$ along $x$-direction and between $0.01r_y$ and $0.7r_y$ spaced by $0.0004r_y$ along $y$-direction). These $2500$ test particles has $\omega_x=1.05461$ and $2455$ test particles has $\omega_y=1.05461$ and $45$ test particles has $\omega_y=1.00667$. We used FFTW (The Fastest Fourier Transform in the West)~\cite{FFTW} to compute numerically the frequencies.}
\label{fft}
\end{figure}

The emittance evolution given by equation~(\ref{eq5}) is shown in Fig.~\ref{emit}. The initial transversal emittance is equal $\epsilon_{x0}=\epsilon_{y0}=0.22360$~\cite{Simeoni2}. It is observed emittance coupling caused for space charge driven nonlinear difference single particle resonance~\cite{Montague1,Hofmann5,Ohnuma1}. This coupling is characterized by the emittance exchange between the directions. Space-charge alters the net force seen by the individual particles in a way that is nonlinear and dependent on the density distribution of the beam. Emittance exchange requires resonant coupling, which can take place only if an intrinsic resonace relationship is fulfilled. A simplified approach would be a difference single particle resonance condition $l\omega_x-m\omega_y=0$ which was suggested in Ref.~\cite{Lagniel3}. $\omega_x$ is single particle dimensionless frequencies along of $x$-direction and $\omega_y$ is single particle dimensionless frequencies along of $y$-direction, $l$ and $m$ are integer numbers. The motion is always bounded in both directions . This emittance coupling is a manifestation of the exchange of energy from one to the other directions which is familiar in the linear coupling $l=m=1$. We used FFTW (The Fastest Fourier Transform in the West) to compute numerically the frequencies $\omega_x$ and $\omega_y$ of $2500$ test particles. We launch the test particles along the $x$- and $y$-axes of beam in specified region inside the beam (between $0.01r_x$ and $0.7r_x$ spaced by $0.0004r_x$ along $x$-direction and between $0.01r_y$ and $0.7r_y$ spaced by $0.0004r_y$ along $y$-direction). These $2500$ test particles has $\omega_x=1.05461$ and $2455$ test particles has $\omega_y=1.05461$. Thus $2455$ test particles obey resonance condition $\omega_x-\omega_y=0$. The results are presented by the histogram in Fig.~\ref{fft}. This number large of test particles in resonance cause emittance coupling. The emittance growth is induced by nonlinear space-charge forces. $\epsilon_x$ increases and $\epsilon_y$ decreases with increasing $r_x/r_y$ but $\epsilon_y$ increases and $\epsilon_x$ decreases with decreasing $r_x/r_y$ as observed in Fig.~\ref{emit} (bottom). The emittance growth is larger in the plane that has the larger semi-axes length. That the emittance growth increases as the semiaxes length increases may seem surprising for an effect that arises from space-charge force, which increases as the beam size becomes smaller, rather than larger and vice versa. $(\epsilon_x+\epsilon_y)/2$ shows a much smoother but similarly pronounced anisotropic variable $\chi$ response.

\begin{figure}[htb]
\centering
\includegraphics*[width=85mm]{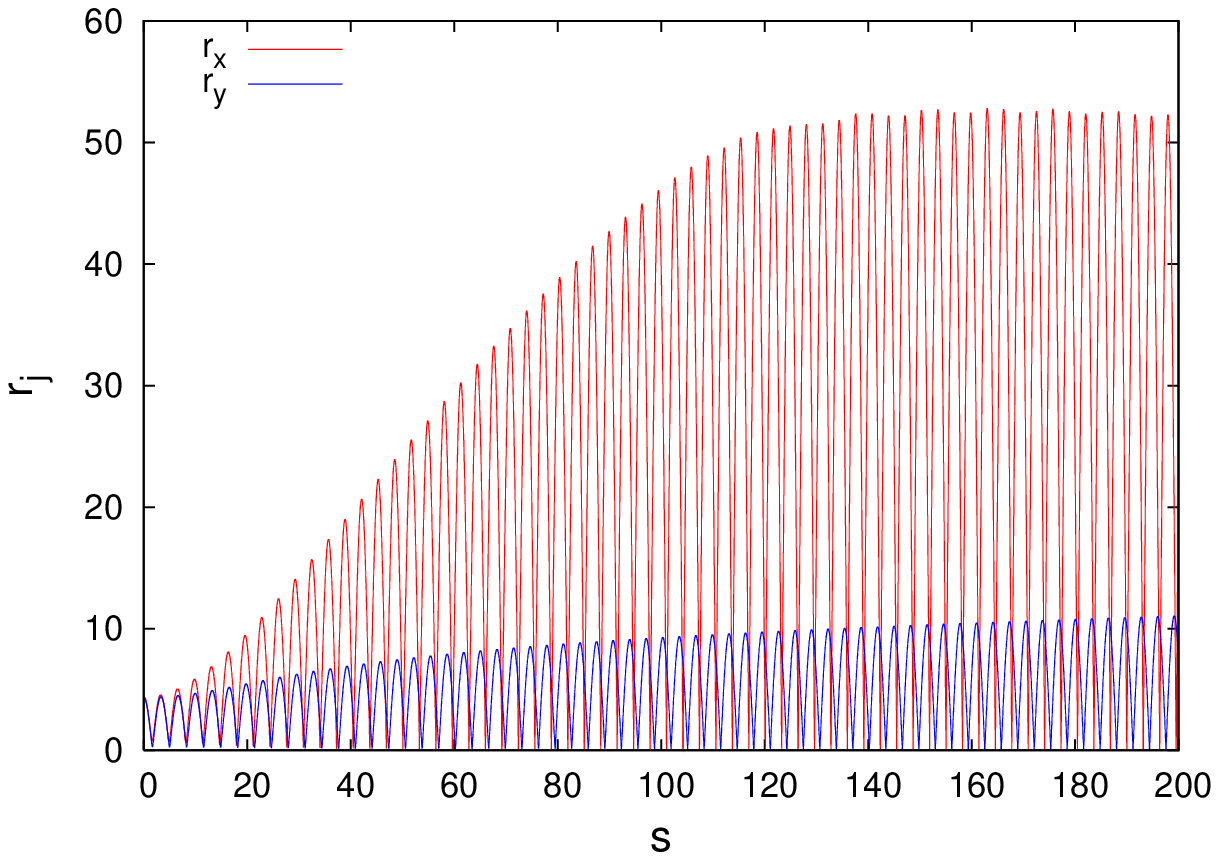} \\
\includegraphics*[width=85mm]{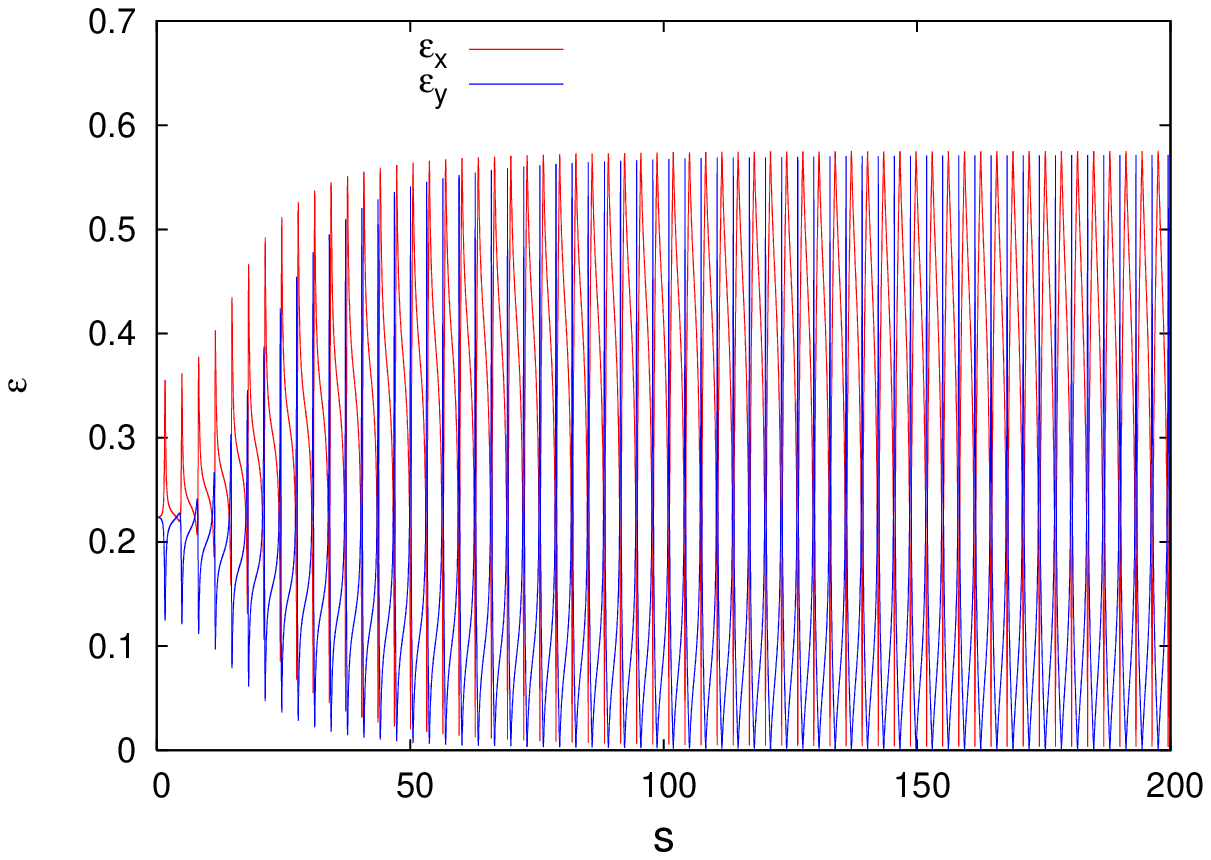}
\caption{Asymtotic evolution of the envelopes (top) and emittance (bottom) obtained of the equations~(\ref{eq4}) and~(\ref{eq5}), respectively. $r_x$~(top),~$\epsilon_x$~(bottom) and $r_y$~(top)~$\epsilon_y$~(bottom) are represented by red and blue lines, respectively.}
\label{asymptotic}
\end{figure}

An important effect of space charge is the tendency to induce waves the beam, a collective effect. These waves are characterized by the plasma frequency, which in turn relates to the perveance $K$ defined as the ratio of the average space charge force to the external focusing force at the beam edge. The effect these waves is observed in the asymtotic evolution of the envelope and emittance shown in Fig.~\ref{asymptotic}. The exponential growth of the envelope and the emittance exchange is characteristic by instability tilting mode of a space-charge-dominated regime beam~\cite{Hofmann2,Hofmann6}. Space-charge in connection with nonlinear dynamics lead to this phenomenon. The parabolic distribution is characterized by the appearence of the tilting mode that emittances are periodically exchanged between $x$ and $y$, similar to a second order difference resonance driven by skew quadrupoles. We note that this second order difference resonance is entirely absent, if the initial emittances are chosen equal. This tilting instability between $x$ and $y$ obviously requires a sufficiently anisotropy. Note that according to~\cite{Simeoni1} the symmetric perturbation only excites the breathing mode, whereas the antisymmetric perturbation excites the quadrupole mode only. With anisotropy this is not the case and we find mixing of breathing and quadrupole modes. In our model both modes are resonance. Space-charge induces a coherent shift of the resonance conditions $l\omega_x\pm m\omega_y=0$, since the full ensemble of particles respond to the resonance in a coherent way. For such a coherently oscillating beam along an uniform linear focusing channel the resonance condition is $l\omega_x\pm m\omega_y+\Delta\omega=0$~\cite{Hofmann7}. For the tilting mode we expect different shifts for the cases $\omega_x+\omega_y+\Delta\omega=0$ (sum resonance) and $\omega_x-\omega_y+\Delta\omega=0$ (difference resonance)~\cite{Hofmann8}. $\Delta\omega$ is the coherent shift away from the single particle resonance condition caused by the coherent motion of all, or a large fraction of particles of beam. It makes the majority of $2500$ test particles launched along the $x$- and $y$-axes inside of the beam to be resonance as shown by the histogram in Fig.~\ref{fft}. In our model the driving term for this difference resonance is not a skew quadrupole as in synchrotrons, but the internal space-charge force caused by the exponentially growing tilting of the beam cross section. There is a strong dependence of the difference resonance on the emittance ratio. In cases where the difference ressonance is crossed for unequal emittance we expect an asymmetric behavior of beam.

Nonlinear resonances eventually yield {\it rms} emittance growth as more and more particles to be launched out of core. Therefore, the halo is not caused by a collective effect involving all core particles but is caused by resonant interaction of some particles with the mismatch modes. Assuming which beam is usually equipartitioned in $x$ and $y$ ($\xi=(r_y\epsilon_x)^2/(r_x\epsilon_y)^2$), where $\xi$ is the ratio of oscillations energies in the $x$ and $y$ directions~\cite{Lloyd1}, but it has large beam size-rms mismatched, the resonances enable ``excess'' energy transfer from one plane to another~\cite{Hofmann1}. We show that the exchange is accompanied by halo creation along one direction preferential.

\begin{figure}[htb]
\centering
\includegraphics*[width=85mm]{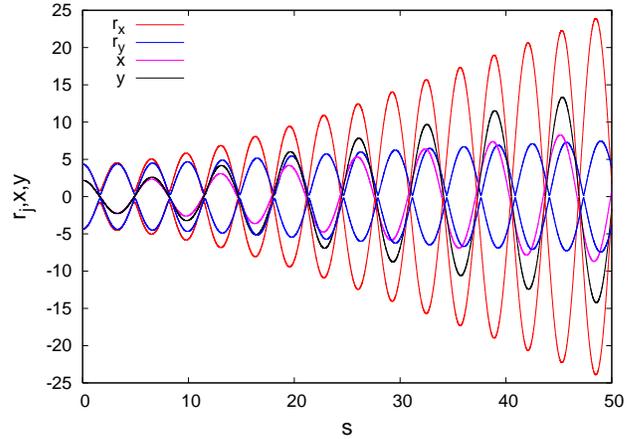}
\caption{Evolution of the dynamical test particle. $x=0.5r_x$ and $y=0.5r_y$, initial.The orbits $x$ and $y$ of test particle are represented by black and pink lines, respectively. The beam envelopes, $r_x$ and $r_y$ are represented by red and blue lines, respectively.}
\label{partcore}
\end{figure}

To analyse the resonances we computed dimensionless frequencies of breathing and quadrupole modes and the dimensionless frequencies $\omega_x$ and $\omega_y$ of $2500$ test particles launched along the $x$- and $y$-axes inside of the beam using Fourier analyses. We used FFTW (The Fastest Fourier Transform in the West) to compute numerically the frequencies. Breathing mode frequency is $\omega_{X_s}=2.01334$ and quadrupole mode frequency is $\omega_{X_a}=2.01334$. Note that both modes has the same frequency. Therefore breathing mode and quadrupole mode are resonance $(1:1)$. $2500$ test particles has $\omega_x=1.05461$ and $2455$ test particles has $\omega_y=1.05461$. Thus $2455$ test particles obey resonance condition $\omega_x-\omega_y=0$, and $45$ test particles has $\omega_y=1.00667$ as shown by the histogram in Fig.~\ref{fft}. The system is dominated by the ressonance $\omega_x-\omega_y=0$, which creates a barrier between the region inside the beam and the region outside the beam~\cite{Cappi1}. The space-charge coupling force is responsible for the energy transfer from the core oscillations to the single particle oscillations. The dominant $1th$ order ressonance at $\omega_x/\omega_y=1$ yields significant emittance exchange~\cite{Hofmann9}. The particle-core (2:1) resonances are present in both quadrupole mode and breathing mode of core oscillation, but only in $y$-direction of test particle~\cite{Ikegami1}. Thus, halo is formed along $y$-direction as illustrated in the Fig.~\ref{partcore}~\cite{Simeoni2}. The key to interpreting anisotropic halo growth due to the $(2:1)$ is the dependence of space-charge force, which increases as the beam size becomes smaller. The parametric $(2:1)$ resonance halo, modified by driving term for the difference resonance that is the internal space-charge force caused by the exponentially growing tilting of the beam cross section, remains the dominat mechanism to explain halo formation along one direction preferential.

\begin{figure}[htb]
\centering
\includegraphics*[width=85mm]{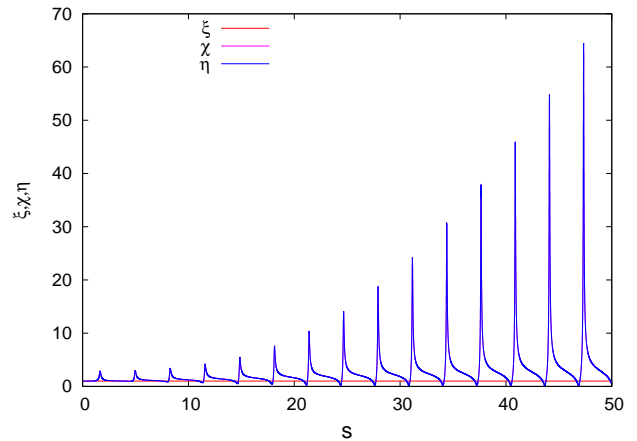} 
\caption{The ratio of oscillations energies $\xi$ and anisotropic ratios, $\chi$ and $\eta$ in beam propagation. $\xi$ is represented by red line, and $\chi$ and $\eta$ are overlapping represented by blue line.}
\label{aniso}
\end{figure}

For large beam size-rms mismatched and initial ratio envelopes beam $\chi = 1$, the ratio of oscillations energies in the $x$ and $y$ directions remaines constant, $\xi=1$~\cite{Jameson2}. As illustrated in the Fig.~\ref{aniso} beam fast suffers anisotropization characterized for discontinuous variations in $\chi$ and $\eta$~\cite{Startsev2,Simeoni2}. The anisotropy leading to coupling resonance~\cite{Hofmann3,Hofmann4} in the presence of nonlinear space-charge forces was suggested as a possible approach to the equipartitioning question~\cite{Wangler2,Lagniel3,Kishek1,Kishek2}, since collisions cannot be made responsible for energy transfer in linacs. The underlying mechanism is collective oscillations of the space-charge density that creates nonlinear forces similar to those magnetic sextupoles and leads to the resonat coupling~\cite{Kishek2,Lund2,Kandrup1}. 

Statistical theories of chaotic systems are generally based on Liouville's theorem and the assumption of equipartition, subject to the constraints given by invariants of the system. If the only invariants of the system are the total energy and the total number of particles, the assumption of equipartition leads to the theory of statistical mechanics at thermal equilibrium. The temperature is then uniform. In turbulence, the energy is not conserved, since there is an external energy source, but there may be other invariants that survice the turbulence. The random flutuations then drive the system toward turbulent equipartition (TEP) on the hypersurface in phase space that is defined by these invariants~\cite{Yankov2}. Previous examples of TEP in plasmas physics were considered by Yankov~\cite{Yankov1} and Isichenko {\it et al.}~\cite{Isichenko1}. Turbulent equipartition corresponds to a phase space density uniform on the surfaces of the two constant adiabatic invariants, a version of the ergodic hypothesis where adiabatics invariants play the role of the no longer conserved energy. The applicability of the statistical mechanics to self-consistent plasma turbulence is not clear, but some predictions of this approach are in a qualitative agreement with experiments~\cite{Yankov1}. In the state of turbulent equipartition, the plasma density and temperature are inhomogeneous, and there are no particle or energy fluxes. Such a state is usually marginally stable to the modes potentially responsible for driving the plasma to this state. We believe that the problem of anisotropic equipartition in linac should be solved in the same spirit.

\section{\label{Equiparti}Anisotropic Kinetic and Dynamic Processes in Equipartitioned Beams}

In order to understand the initial dynamical behavior of an anisotropic beams, in particular, to study possible mechanisms of equipartition connected with phase space we have to know how we can compute the variables (volume, area of surface, and area projected) that characterize the anisotropic beam in phase space. The equipartition is generally based on the assumption that system is ergodic. If the dynamics is chaotic in some subspaces of the phase space, the system can be considered as ergodic in these subspaces. The beam began in a nonequilibrium state evolved toward a metaequilibrium in which the particle orbits filled an invariant measure of phase space. The transient dynamics reflected an intricated network of space-charge modes. We shall call this subspace of the phase space \textquotedblleft~$\xi$~\textquotedblright, where $\xi$ is the ratio of oscillations energies in the $x$ and $y$ directions. Unlike the thermodynamic equipartition energy in conservative systems, anisotropic equipartitions  describe strongly non-equilibrium systems such as high-intensity ions beam in space-charge dominated regime with large mismatched beam size-rms initial. Anisotropic equipartition corresponds to a phase space density uniform on the surface invariant of the ($\xi=(r_y\epsilon_x)^2/(r_x\epsilon_y)^2$), a version of the ergodic hypothesis where the  $\xi$ invariant play the role of the conserved energy~\cite{Kandrup3}. In the state of anisotropic equipartition, the Lyapunov function~\cite{Lyapunov1} and temperature are stationaries, the entropy grows in the cascade form, there is a coupling of transversal emittance, the beam develops an elliptical shape with a increase in its size along one direction and there is halo formation along one direction preferential. Such a state is usually marginally stable.

Emittance is the area in phase space occupied by the particles of the beam. The maximum emittance of a beam that a system can accept is called the acceptance of the system~\textquotedblleft~$a$~\textquotedblright~\cite{Weiss1,Joh1,Hofmann10}. Then the acceptance is defined as the maximum phase space area where particles can survive in the accelerator. It can be determined by dynamical effects associated to space-charge forces. In a linear accelerator with nonlinearities, the transversals motions are coupled. The particles move on the distorced surface in four-dimensional transverse phase space, the so-called \textquotedblleft hyper-egg\textquotedblright . We can talk only about the projections of this hyper-egg onto the two transverse phase planes $a_x(x,x^{'})$ and $a_y(y,y^{'})$. These projections are no longer clean curves but bands. It is well-know fact that in presence of space-charge the beam phase space structure gets quite complicated and is divided to areas of stable and unstable motion. 

Qualitatively, emittance can be thought of as the temperature of the beam, a measure of the random disorder in the transverse momenta of the constituent particles. Quantitatively, transverse emittance is defined by drawing a contour around a given percentage of particles in phase space and gives us a numerical figure of merit for describing the quality of the beam. As the beam is focused or defocused, the convergence or divergence of the beam envelope yields a correlation between particle position and angle of motion; if, however, we focus the beam to a waist, the correlation between particle position and transverse momentum
is minimized. We define the RMS acceptance as the maximum square root of the $a_j=\sqrt{\langle j^{2}\rangle \langle j^{'2}\rangle -\langle jj^{'}\rangle ^2}$ (henceforth, $j$ ranges over both $x$ and $y$, where $x$ and $y$ are the positions of the beam particles). The second moment term $\langle jj^{'}\rangle ^2$ represents a correlation between $j$ and $j^{'}$ that exists when the beam envelope is converging or diverging; qualitatively, it can be thought of as a measure of inward or outward flow of transverse kinetic energy. At a waist, this correlation is minimized and the second moment term is zero~\cite{Reiser1}. Thus, the acceptance reduces to maximum of the $\tilde{a}_j=\sqrt{\langle j^{2}\rangle \langle j^{'2}\rangle}$. It is a very important for one to know how much of the phase space is stable. With the help of the Lyapunov functions we can construct an invariant beam area. This function is dependent of the equipartition $\xi$, of the anisotropy variable $\eta$, ratio of the average external focusing force to the space-charge force $\alpha$ and space-charge strength $\rho$.

It is intuitively clear that if the total energy of a physical system has a local minimum at a certain equilibrium point, then that point is stable. This idea was generalized by Lyapunov~\cite{Lyapunov1} into a simple but powerful method for studying stability problems in a broader context. Lyapunov functions are functions which can be used to prove the stability of a certain fixed point in a dynamical system or autonomous differential equation. For dynamical systems (e.g. physical systems), conservation laws can often be used to construct a Lyapunov function. A certain subclass of dynamical systems, namely potential or gradient system are of particular interest because their behaviour is simpler than the general case, and because they are frequently encountered in approximate treatments of physical systems. For a gradient system, $\frac{d x(t)}{d t} = f(x)$, takes the form $ \frac{d x(t)}{d t} = -\frac{d L(x)}{d x}$, where  $L(x)$ is a function of the variable $x$. More generally, a system with an equilibrium $x_0$ is said to have a Lyapunov function $L$ for this equilibrium if this function satisfies the conditions $L(x_0)=0$, $\frac{d L(x)}{d t}\leq 0$ for $x\neq x_0$ and $L$ is a smooth function of $x$ in some neighborhood of $x_0$~\cite{Hirsch1}. A gradient system, satisfying $ \frac{d x(t)}{d t} = -\frac{d L(x)}{d x}$ has global Lyapunov function if $L$ is bounded. For gradient system the dynamic consists of relaxation toward minimum in $L$. This means that such functions are strictly only defined when corresponding equilibriums are fixed points. Finding a Lyapunov function for a certain equilibrium might be a matter of luck. Trial and error is the method to apply, when testing Lyapunov functions on some equilibrium.

\begin{figure}[htb]
\centering
\includegraphics*[width=85mm]{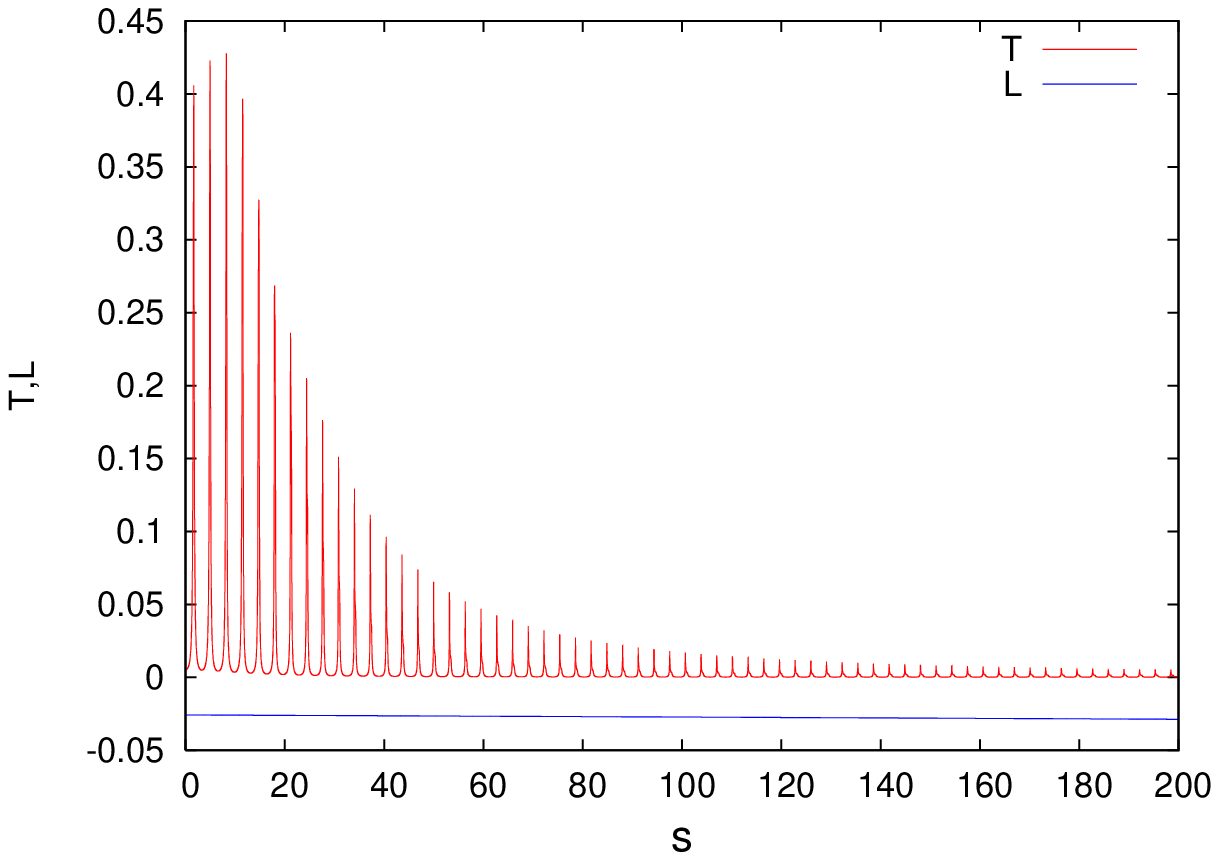} \\
\includegraphics*[width=85mm]{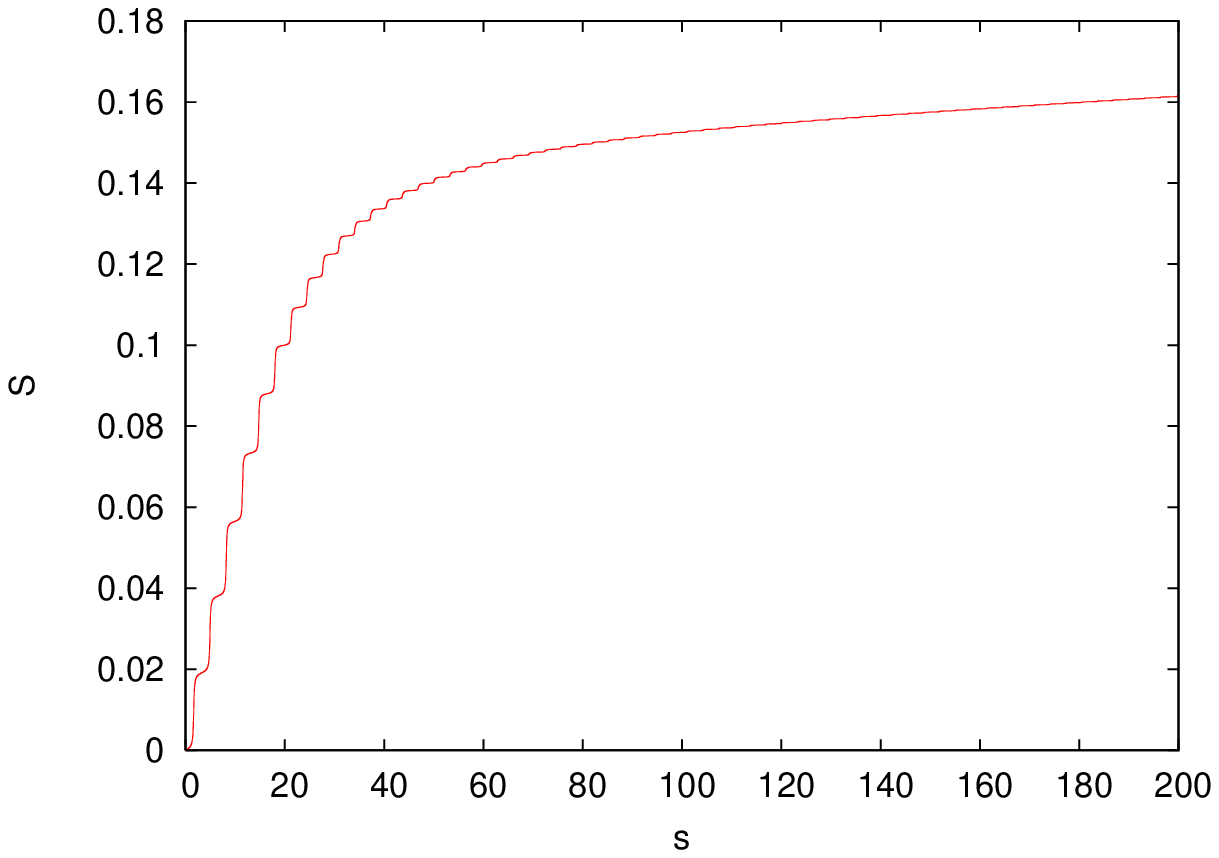} 
\caption{Evolutions of the  temperature $T_{eq}$ (top), Lyapunov function $L$ (top) and entropy $S$~(bottom) obtained of the equations~(\ref{11}),~(\ref{eq6}) and (\ref{12}), respectively. $T_{eq}$~(top) and $S$~(bottom) are represented by red lines and $L$~(top) is represented by blue line.}
\label{lyapu}
\end{figure}

Next, we will apply Lyapunov's method above to the acceptance dynamics~\cite{Frank1}. To this end, we will replace the aforementioned quantities $x(t)$ and $L(x)$ by the $\tilde{a}_j(s)$ acceptance, and $L(\tilde{a}_j)$ Lyapunov function of the acceptance. A definition of a Lyapunov function would be : 

\begin{eqnarray}
\label{eq6}
L&=&\alpha\left(1-\sqrt{1/\xi}\right)\frac{\tilde{a}^2_x}{2r_x}-\rho \left(1+1/\eta \right)\frac{\tilde{a}^2_x}{2} \\ \nonumber 
&-&
\alpha\left(\sqrt{\xi}-1\right)\frac{\tilde{a}^2_y}{2r_y}-\rho \left(1+\eta \right)\frac{\tilde{a}^2_y}{2}
\end{eqnarray}
where $\tilde{a}_j$ (henceforth, $j$ ranges over both $x$ and $y$, where $x$ and $y$ are the positions of the beam particles) is acceptance in the absence of directional correlations between $j$ and $j^{'}$, primes denote derivates with respect to $s$. This function is dependent of the beam envelope $r_j$, of the equipartition $\xi$, of the anisotropy variable $\eta$, ratio of the average external focusing force to the space-charge force $\alpha$ and space-charge strength $\rho$. To $\tilde{a}_{x_0}=0$ and $\tilde{a}_{y_0}=0$, $L$ satisfies the condition $L(\tilde{a}_{x_0},\tilde{a}_{y_0})=0$. It can be interpreted as the equation of a surface that resembles a paraboloid opening downward and tangent to the $a_xa_y$ plane at the origin. $L$ can be considered as a \textquotedblleft free energy \textquotedblright~\cite{Frank1}. To the issue $\frac{d L(\tilde{a}_j)}{d s}\leq 0$ mentioned above the evolution of the $L(\tilde{a}_j)$ Lyapunov function represented by blue line (top graphic) in the Fig.~\ref{lyapu} can be regarded as a proof. As the Lyapunov function~(\ref{eq6}) satisfies the conditions $L(\tilde{a}_{j_0})=0$ and $\frac{d L(\tilde{a}_j)}{d s}\leq 0$ we can apply equation $ \frac{d \tilde{a}_j(s)}{d s} = -\frac{d L(\tilde{a}_j)}{d \tilde{a}_j}$ obtaining the following acceptance dynamics equations: 

\begin{eqnarray}
\label{eq7a}
\frac{d\tilde{a}_x}{ds} &=& -\alpha \left(1-\sqrt{1/\xi}\right)\frac{\tilde{a}_x}{r_x}+ \rho \left(1+1/\eta \right)\tilde{a}_x \\ \label{eq7b}
\frac{d\tilde{a}_y}{ds} &=& \alpha \left(\sqrt{\xi}-1\right)\frac{\tilde{a}_y}{r_y}+ \rho \left(1+\eta \right)\tilde{a}_y 
\end{eqnarray}
where $s$ is the axial coordinate of a beam. Equations~(\ref{eq7a}) and~(\ref{eq7b}) express the acceptance dynamics in term of the beam envelope $r_j$, of the equipartition $\xi$, of the anisotropy variable $\eta$, ratio of the average external focusing force to the space-charge force $\alpha$ and space-charge strength $\rho$. We derive equations for acceptance change in each plane for continuous elliptical beam within a constant focusing channel without correlation between particle position and transverse momentum. The equation will apply to beams within a conducting pipe whose radius is much larger than the beam size. The resulting acceptance equations contain two terms: the first term describe acceptance changes associated with transfer of energy between the two planes; the second describes acceptance changes associated with anisotropic processes. The anisotropy leads to coupling resonance~\cite{Hofmann3,Hofmann4,Wangler2} in the presence of nonlinear space-charge forces. Space-charge couple some degrees of freedom causing resonance between them. Coupling resonance leading to an exchange of energy between the degrees of freedom.

The general solutions to the equations~(\ref{eq7a}) and~(\ref{eq7b}) are given by:

\begin{eqnarray}
\label{eq9a}
\tilde{a}_x &=& c_1e^{\int\left[ \frac{\alpha}{r_x}\left(\sqrt{1/\xi}-1\right)+\rho\left(1+1/\eta\right)\right]ds} , \\ \label{eq9b}
\tilde{a}_y &=& c_2e^{\int\left[ \frac{\alpha}{r_y}\left(\sqrt{\xi}-1\right)+\rho\left(1+\eta\right)\right]ds}
\end{eqnarray}
where $c_1$ and $c_2$ are arbitrary constants determined by initial values of the solutions~(\ref{eq9a}) and~(\ref{eq9b}), respectively.

The oscillations beam envelope perturb nonlinear space-charge force yields a correlation between particle position and transverse momentum. As the beam compresses and expands the particles gaining and losing kinetic energy. Thus, the second moment term $\langle jj^{'}\rangle ^2$ is not zero and the RMS acceptance becomes $a_j=\sqrt{\langle j^{2}\rangle \langle j^{'2}\rangle -\langle jj^{'}\rangle ^2}$. Therefore the equipartition and the variable anisotropy are given by $\xi=(r_ya_x)^2/(r_xa_y)^2$ and $\eta=a_x/a_y$, respectively. Thus the equations~(\ref{eq7a}) and~(\ref{eq7b}) are transformed by :

\begin{eqnarray}
\label{eq8a}
\centering
\frac{da_x}{ds} &=& \left( \frac{-\alpha}{r_x} + \rho \right)a_x + \left( \frac{\alpha}{r_y} + \rho \right)a_y , \\ \label{eq8b}
\frac{da_y}{ds} &=& \left( \frac{\alpha}{r_x} + \rho \right)a_x + \left( \frac{-\alpha}{r_y} + \rho \right)a_y
\end{eqnarray}
where the terms $a_x/r_x$ and $a_y/r_y$ model the equipartitoning and the term $a_x+a_y$ covers residual growth from nonlinear resonance~\cite{Ohnuma1}. Jameson has derived identical equations~\cite{Jameson2,Jameson3,Jameson4} to the question of equipartioning in linear accelerator.

To solve the equations~(\ref{eq8a}) and~(\ref{eq8b}) we add ~(\ref{eq8a}) and~(\ref{eq8b}) and then we solve for $a_x + a_y= c_3e^{(2\rho s)}$. Then it is easiest to substract ~(\ref{eq8b}) from ~(\ref{eq8a}) and substitute for $a_x=-a_y + c_3e^{(2\rho s)}$ to obtain the general solutions to the equations~(\ref{eq8a}) and~(\ref{eq8b}) given by:
\begin{widetext}
\begin{eqnarray}
\label{eq10a}
a_x &=& c_3e^{2\rho s}-\left\{ c_3\int \left[ \frac{\left(e^{2\rho s + \alpha \int \left( -\frac{1}{\chi r_y}+\frac{\chi}{r_x}\right)ds}\right)\left(\alpha+\rho r_x\right) }{r_x} \right]ds+c_4\right\} \left\{e^{\alpha\int\left(\frac{1}{\chi r_y}-\frac{\chi}{r_x}\right)ds}\right\} , \\ \label{eq10b}
a_y &=& \left\{ c_3 \int \left[ \frac{\left(e^{2\rho s + \alpha \int \left( -\frac{1}{\chi r_y}+\frac{\chi}{r_x}\right)ds}\right)\left(\alpha+\rho r_x\right) }{r_x}\right]ds+c_4\right\} \left\{e^{\alpha\int\left(\frac{1}{\chi r_y}-\frac{\chi}{r_x}\right)ds}\right\}
\end{eqnarray}
\end{widetext}
where $c_3$ and $c_4$ are arbitrary constants determined by initial values of the solutions~(\ref{eq10a}) and~(\ref{eq10b}), respectively. $a_j$ is dependent of the beam envelope dynamic $r_j$, of the anisotropy variable variations $\chi=r_x/r_y$ and space-charge effects, $\alpha$ and $\rho$.

To analise the effect of coupling between the two transverse phase planes without ($\tilde{a}_x(x,x^{'})$ and $\tilde{a}_y(y,y^{'})$), and with ($a_x(x,x^{'})$ and $a_y(y,y^{'})$) correlation between particle position and transverse momentum (the second moment term $\langle jj^{'}\rangle ^2=0$ and $\langle jj^{'}\rangle ^2\neq0$, respectively) we solve the integrals in~(\ref{eq9a}) and~(\ref{eq9b}),~(\ref{eq10a}) and~(\ref{eq10b}) to obtain the acceptance evolution without and with correlation, respectively. We launch the beam with $\alpha=0.00001$ and $\rho=0.25$ in space-charge dominated regime, meaning that the collective oscillations dominate over the individual particles betatron motion. $c_1=0.22360$ and $c_2=0.22360$, $c_3=0.44721$ and $c_4=0.22360$ are determined by initial values $\tilde{a}_{x0}=\tilde{a}_{y0}=0.22360$, and $a_{x0}=a_{y0}=0.22360$ of the solutions~(\ref{eq9a}) and~(\ref{eq9b}), and~(\ref{eq10a}) and~(\ref{eq10b}), respectively. We solve the integrals in~(\ref{eq9a}) and~(\ref{eq9b}),~(\ref{eq10a}) and~(\ref{eq10b}) up to $s=50.0$. The corresponding evolution of the $rms$-emittance~(\ref{eq5}) and the $rms$-envelope~(\ref{eq4}) are used. 

\begin{figure}[htb]
\centering
\includegraphics*[width=85mm]{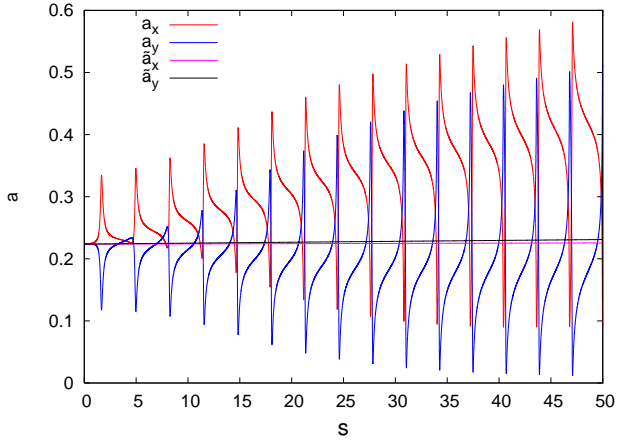} 
\caption{Evolution of the beam acceptance without ($\tilde{a}_x$ and $\tilde{a}_y$), and with ($a_x$ and $a_y$) correlation between particle position and transverse momentum. $\tilde{a}_x$ and $\tilde{a}_y$ are represented by pink and black lines, respectively. $a_x$ and $a_y$ are represented by red and blue lines, respectively.}
\label{accept}
\end{figure}

The acceptance evolutions without and with correlation between particle position and transverse momentum given by equation~(\ref{eq9a}) and~(\ref{eq9b}),~(\ref{eq10a}) and~(\ref{eq10b}), respectively are shown in Fig.~\ref{accept}. The area in $x-x^{'}$ and $y-y^{'}$ trace spaces defined by $\tilde{a}_x$~(pink lines) and $\tilde{a}_y$~(black lines), respectively, remain constant with very small oscillations because $\tilde{a}_x$ and $\tilde{a}_y$ are beam area projections derived from Lyapunov functions~(\ref{eq6}). Also, if a energy-balanced beam is injected, the distribution will move toward equipartition, in which the beam assumes a uniform distribution on the surface of $\tilde{a}_x$ and $\tilde{a}_y$~\cite{Yankov1}. However, it is observed transverse phase planes coupling in $x-x^{'}$ and $y-y^{'}$ trace spaces defined by $a_x$~(red lines) and $a_y$~(blue lines) caused for space charge driven core-core resonance together with single-particle resonances~\cite{Montague1,Hofmann5,Ohnuma1}.This coupling is characterized by the energy exchange between the directions. Energy/acceptance exchange requires resonant coupling, which can take place only if an intrinsic resonance relationship is fulfilled. A significant number of particles are trapped inside a resonance island~\cite{Cappi1,Franchetti1} as shown in Fig~\ref{fft}. They are characterizing the formation of a metastable state of the particles beam~\cite{Yankov1, Satogata1}. Geometrical methods can be used to analyze this state~\cite{Jameson5,Pettini1}. By definition of acceptance we observe the similarity with the evolution of emittance~($\epsilon_x$ and $\epsilon_y$) shown in Fig.~\ref{emit}.

In the process of acceptance transfer, strong correlations between particle position and transverse momentum are naturally developed. The acceptance oscillations are now driven by variations of the space-charge force due to the beam compresses and expands. The rapid aceptance oscillations are due to coherent transverse plasma oscillations in the beam and are a manifestation of periodic energy exchange between potential and kinetic energies. The acceptance growth results from nonlinearities in the particles’ oscillations about their equilibrium. Excess energy is required for driving the nonlinearities, which stems from the energy anisotropy between different degrees of freedom. Acceptance grows in a plane that receives energy, and decreases in a plane that loses energy. The acceptance $a_x$ growth increased, and the acceptance $a_y$ growth decreased and vice-versa during the propagation of the beam. Such transfers between $x-x^{'}$ and $y-y^{'}$ are commonly observed if the initial acceptance differ. This indicates that ratios acceptance are important variables. In order to expedite an acceptance flow from one direction to another, the two directions have to be on coupling resonance. Typically, we employ difference resonance: $l\omega_x-m\omega_y=0$ where $\omega_x$ is single particle frequencies along of $x$-direction and $\omega_y$ is single particle frequencies along of $y$-direction, $l$ and $m$ are integer numbers. In our case $l=m=1$ and number large of particles obey resonance condition $\omega_x-\omega_y=0$ as shown in Fig.~\ref{fft}. Acceptances only start to flow from a \textquotedblleft hot  \textquotedblright to a \textquotedblleft cold \textquotedblright degree of freedom. This suggests the development of correlation among the degrees of freedom during the acceptance exchanging process.

Energy/acceptance flow is due to the correlation $jj^{'}$ between the velocity and the position of the particle in regions where the beam size contracts or expands. The term $\langle jj^{'}\rangle ^2$ represents an inward or outward flow term in the transverse kinetic energy. The physical interpretation is that the rms kinetic enrgy of the particle distribution consists of a thermal component and a flow component~\cite{Reiser1}. Transfer of kinetic energy from one coordinate direction to another results in partial or complete kinetic energy equipartitioning. While superficially similar to equipartitioning of energy in a gas, this transfer does not result from individual particle collisions but, presumably, from interactions between the individual particles and collective fields. Hofmann has show that coherent-mode instabilities can lead to kinetic-energy exchange~\cite{Hofmann3}.

In subsequent analysis, the ratio $a_x/a_y$ is varied to explore different degrees of anisotropy and energy flow. The rate of exchange can be predicted with certainty analytically. To obtain the relation between $\tilde{a}_x$ and $\tilde{a}_y$, and $a_x$ and $a_y$, respectively, which will give the acceptance flow from one direction to another, elimanate $s$ by dividing~(\ref{eq7a}) by~(\ref{eq7b}), and ~(\ref{eq8a}) by~(\ref{eq8b}), respectively. The equations for the acceptance flow are:

\begin{eqnarray}
\label{eq15a}
\centering
\frac{d\tilde{a}_x}{d\tilde{a}_y} &=& \frac{\left[\frac{-\alpha \left(1-\sqrt{1/\xi}\right)+ \rho \left(1+1/\eta \right)r_x}{r_x}\right]\tilde{a}_x}{\left[\frac{\alpha \left(\sqrt{\xi}-1\right)+ \rho \left(1+\eta \right)r_y}{r_y}\right]\tilde{a}_y}, \\ \label{eq15b}
\frac{da_x}{da_y} &=& \frac{\left(-\alpha+\rho r_x \right)r_ya_x + \left(\alpha+\rho r_y \right)r_xa_y}{\left(\alpha+\rho r_x \right)r_ya_x + \left(-\alpha+\rho r_y \right)r_xa_y}
\end{eqnarray}

The equations~(\ref{eq15a}) and~(\ref{eq15b}) are separable variables. The general solutions are :

\begin{widetext}
\begin{eqnarray}
\label{12a}
F_{\tilde{a}_y/\tilde{a}_x} &=& \eta \ln\left(a_x\right) - \ln\left(a_y\right) , \\ \label{12b}
F_{a_y/a_x} &=& \frac{b_1\ln\left(b_3\Theta^2+b_4\Theta+b_5\right)}{2b_3}+\left(\frac{2b_2-\frac{b_1b_4}{b_3}}{\sqrt{4b_5b_3-b_4^2}}\right)\arctan\left(\frac{2b_3\Theta+b_4}{\sqrt{4b_5b_3-b_4^2}}\right)-\ln\left(a_x\right)
\end{eqnarray}
\end{widetext}
where $\eta=\epsilon_x/\epsilon_y$, $\Theta=a_x/a_y$, $b_1=-r_y\left(\rho r_x +\alpha\right)$, $b_2=r_x\left(\rho r_y -\alpha\right)$, $b_3=r_y\left(\rho r_x +\alpha\right)$, $b_4=\alpha\left(r_x -r_y\right)$, $b_5=r_x\left(\rho r_y +\alpha\right)$. $r_x,r_y$ is the beam envelope, $\alpha$ is the ratio of the average external focusing force to the space-charge force and $\rho$ is the space-charge strength. $F_{\tilde{a}_y/\tilde{a}_x}$ is the acceptance flow without correlation and $F_{a_y/a_x}$ is the acceptance flow with correlation.

\begin{figure}[htb]
\centering
\includegraphics*[width=85mm]{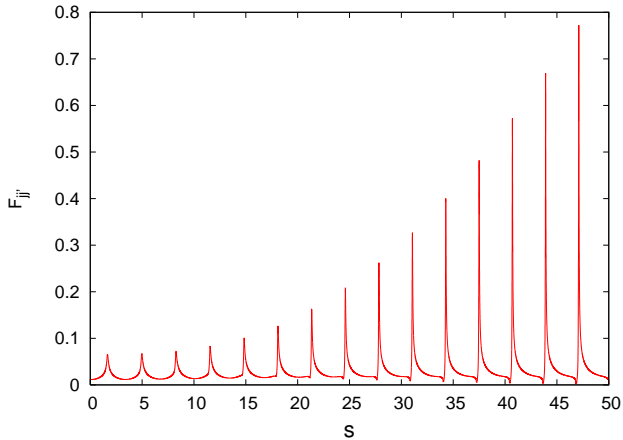} 
\caption{Evolution of the flow component $F_{jj^{'}}$  obtained of the difference between equation~(\ref{12b}) and the equation~(\ref{12a}), $F_{jj^{'}}=F_{a_y/a_x}-F_{\tilde{a}_y/\tilde{a}_x}$ up to $s=50.0$.}
\label{flux}
\end{figure}

The necessary conditions for efficient heat transfer between transverse degrees of freedom are formulated. As discussed, $\langle jj^{'}\rangle ^2$ represents an inward or outward flow term in the tranverse kinetic energy. The tranverse kinetic energy consists of a thermal component and a flow component. The latter is due to the correlation between the velocity and the position of the particle in regions where the beam size contracts or expands. Flow component is then directly related to heat transfers between different degrees of freedom within the beam. The difference between the acceptance flow with correlation and the acceptance flow without correlation is the flow component defined by  $F_{jj^{'}}=F_{a_y/a_x}-F_{\tilde{a}_y/\tilde{a}_x}$. The flow component evolution $F_{jj^{'}}$ is shown in Fig.~\ref{flux}. It is affected by a swift variation similar discontinuous variations in $\chi$ and $\eta$ in Fig~\ref{aniso}. The amplitude oscillations grow during the propagation of the beam. When the beam radius expands, the flow component has an outward direction and the thermal component decreases, when the beam radius contracts, the flow component is inward and the thermal component increases. The propagation of the beam is thus characterized by a variation in the envelope beam which correlates with the generation of the heat flow and a variation of the beam temperature. As usual in statistical physics, we relate the temperature of a particle ensemble to its random motion. Charged particle beams change their size while passing through an ion optical system. The tranverse kinetic energy  $\langle p^2_j\rangle/2m$ contains a flow component if  $\langle jj^{'}\rangle \neq 0$. Therefore, the flow component of the kinetic energy must be subtracted from transverse kinetic energy in order to obtain its thermal component. The \textquotedblleft non-equilibrium temperature \textquotedblright~$T_j$ is defined as the thermal component of the transverse kinetic energy of the {\it j-th} degree of freedom. 
 
The elementary mechanism responsible for the transfer of heat from one degree of freedom to another is constituted by the effect of interactions between the individual particles and waves of collective fields~\cite{Elskens1}. The transfer of heat arises from the scattering of a particles and waves by a mean-background of waves of collective fields in thermal equilibrium via resonant interactions~\cite{Lvov1,Zakharov1}. We demonstrate that resonant interactions of the particles beams provide a mechanism for effective energy exchanges among different degree of freedom. The issue of reaching thermal equilibrium is to imagine a heat bath, which is in thermal contact with the beam. In this case, the particles of the beam can exchange energy through the thermal bath. Chavanis~\cite{Chavanis1} uses a kinetic theory to describe the relaxation of a test particle in a thermal bath of field particles. The relaxation of the system is due to a condition of resonance and it may happen that the relaxation stops because there is no resonance anymore.It has been shown that all heat transfers within the beam — feeding thermal energy from one degree of freedom to another — are always associated with an increase of beam entropy and thus always lead to an irreversible degradation of beam quality. Since the particle distribution of a beam is confined by focusing potentials and the individual particles are performing oscillations, there is a continuous exchange between potential energy and kinetic energy so that displacements in position due to random processes translate into velocity changes and vice-versa. The basic idea is that a continuous exchange between potential energy and kinetic energy can trigger chaos via a parametric resonance~\cite{Kandrup1,Gluckstern1}.

A priori there need be no direct connection between increases in chaos and exchange of energies. However, one would expect resonant couplings to lead significant changes in energies. A correlation between changes in energy and the amount of chaos thus corroborates the interpretation that this chaos is resonant in origin. This transient chaos can drive chaotic phase mixing, which, in the context of a fully self-consistent evolution, might be expected to play an important role in violent relaxation~\cite{Lynden-Bell1,Kull1,Kadomtsev}. After the energy redistributes among all the modes to achieve \textquotedblleft thermal\textquotedblright equilibration~\cite{Chavanis2}. Recent simulations~\cite{Kishek3} have shown that fully self-consistent simulations of beams can exhibit evidence of chaotic phase mixing.

To characterize the \textquotedblleft thermal \textquotedblright equilibrium we analyze the dynamics of the Lyapunov function~(\ref{eq6}), and of the temperature and entropy of the beam. The \textquotedblleft temperature \textquotedblright $T_j$  of the {\it j-th} degree of freedom of a charged particle beam can be expressed in terms of second order beam moments $T_j=\frac{ma^2_j}{\Bbbk_br^2_j}$, where $m$ is the mass ions beam and $\Bbbk_b$ is the Boltzmann's constant~\cite{Toepffer1,Struckmeier2}. This formula shows that envelope $r_j$ and acceptance $a_j$ variations cause temperature variations. For a system with \textquotedblleft non-equilibrium temperatures \textquotedblright that is oscillating anisotropicly around $T_{eq}$, the equilibrium temperature can therefore be aproximated by the arithmetric average of the $T_j$ :

\begin{equation}
\label{11}
T_{eq} = \frac{m}{2\Bbbk_b}\left(\frac{a_x^2}{r_x^2}+\frac{a_y^2}{r_y^2}\right)
\end{equation}

With the equilibrium temperature $T_{eq}$ for the 2-D beam model, the entropy change near thermodynamic equilibrium due to a temperature balancing process may then be written as

\begin{equation}
\label{12}
\frac{dS}{ds} = \frac{\Bbbk_b}{2}\left[\frac{\left(T_x-T_y\right)^2}{T_xT_y}\right] 
\end{equation}

Obviously, the entropy $S(s)$ remains unchanged in the case of temperature equilibrium while increasing during temperature balancing. We may regard Eq.(\ref{12}) as a particular manifestation of Boltzmann's H-theorem~\cite{Tremaine1}. The basis for the dynamics behaviour of the entropy is the relation between the acceptance, the beam temperature and the envelope. Within the beam, heat exchange between the degrees of freedom may occur, leading to an entropy growth as described by Eq.(\ref{12}). We conclude that equipartitioning effects occurring within initially thermally unbalanced charged particle beams are always associated with an irreversible degradation of the beam quality as a whole. Beam transport without an increase of entropy are thus possible if either the beam stays round throughout its propagation. The entropy concept was first applied to beam, and its relationship to {\it rms} emittance was explored by Lawson, Lapostolle and Gluckstern~\cite{Lawson1}. In 1991, the maximum-entropy hypothesis was used to calculate the characteristic of the final distribution for a high-intensity expanding beam in free space~\cite{Connell1}.

We launch the beam with $m=1$, $r_{j0}=0.43616$ and $a_{j0}=0,22360$, consequently $T_{j0}=0.26282$ initially, and integrate the entropy equation~(\ref{12}) up to $s=200.0$. The corresponding evolution of the temperature $T_j$ is used in Eq.(\ref{12}). To the evolution of the temperature $T_j$ the corresponding evolution of the $rms$-acceptance~(\ref{eq8a}) and (\ref{eq8b}), and the $rms$-envelope~(\ref{eq4}) are used. To the Lyapunov function dynamics the corresponding evolution of the $rms$-acceptance~(\ref{eq7a}) and (\ref{eq7b}) and the $rms$-envelope~(\ref{eq4}) are used. The dynamics of the entropy $S$ (bottom), temperature $T_{eq}$ (top) and Lyapunov function $L$ (top) are shown in Fig.~\ref{lyapu}.

In the case without correlation between particle position and transverse momentum the Lyapunov function $L$ is a monotonically decreasing function with respect to $s$ for solutions $\tilde{a}_j$ (because we have $\frac{d L(\tilde{a}_j)}{d s}\leq 0$) as shown in Fig.~\ref{lyapu}; the fixed points $\tilde{a}_{x_0}=0$ and $\tilde{a}_{y_0}=0$ correspond to extrema of $L$ (i.e. we have $ \frac{d \tilde{a}_j(s)}{d s}=0 \Leftrightarrow \frac{d L(\tilde{a}_j)}{d \tilde{a}_j}=0$) and as $L$ is bounded then from the two aforementioned properties it follows that $L$ becomes stationary in the limit $s\rightarrow\infty$. The stationarity of the Lyapunov function $L$ implies the stationarity of the ions beam.

In the case with correlation between particle position and transverse momentum the temperature $T_{eq}$ oscillates after steady with small fluctuations and entropy $S$ grows by cascades as shown in Fig.~\ref{lyapu}. By analogy with compression and expansion of a gas, the beam temperatute $T_{eq}$ heats up during compression and cools during expansion. The concept of entropy cascade is the key agent in the heating and relaxation of the beam~\cite{Howes1}.The physical process by which this relaxation occur is the wave-particle interaction rather than the familiar Coulomb collisions: the reservoir of free energy, represented by the $T_x/T_y > 1$ or $T_x/T_y < 1$  states, pump energy into initially small microscopic fluctuations in the system and lead to the build up of a significant level of electromagnetic anisotropic fields. This is done initially in the temperature $T_{eq}$  which therefore decreases in time. As a non-linear feedback effect, the waves heat the particles of the system. The feedback effect of the waves on the particles results in a final state in which particles and fields are in a quasi-stationary state~\cite{Ruffo1,Chavanis3}. Qualitatively, the collisionless relaxation is driven by the fluctuations of the field, itself induced by the fluctuations of distribution function. The fluctuations of the electromagnetic fields are able to redistribute energy between particles beam and provide an effective relaxation mechanism on a very short timescale. Physically, this collisionless relaxation is interpreted as reflecting a resonant coupling of particles with the wave of fields~\cite{Kadomtsev}. The resulting \textquotedblleft resonant phase mixing \textquotedblright might be sufficiently strong to explain violent relaxation. During violent relaxation, the beam tends to maximize the rate of entropy production $\frac{dS}{ds}$ while conserving the constraints imposed by the dynamics. Moreover, as the system approaches quasi-equilibrium, the fluctuations of the field are less and less efficient.

Collisionless Landau damping of the electromagnetic fluctuations leads to particle heating in the sense that it transfers the electromagnetic fluctuation energy into fluctuations of the particle distribution function, which are then converted into heat by stochastic space-charge forces~\cite{Brown1}. The nonlinear collective forces have the same effect as collisions in thermalizing a particle distribution. The available energy is, however, not entirely thermalized. Some of the energy will be converted to potential energy, due to the change in beam radius. Stochastic space-charge forces are required to increase the entropy. The entropy cascade is the way in which the energy diverted from the electromagnetic fluctuations by the collisionless damping (wave-particle interaction) can be transferred to the stochastic space-charge forces scales resulting in equipartitioning. Equipartitioning via this anisotropic instabilities can properly be described in terms of a thermalization of the beam in the sense of diffusion~\cite{Bohn2,Chavanis4,Tzenov1}. The beam distribution never reaches the true equilibrium distribution. Instead, two distinct distributions are seen the majority of the beam is very nearly in equilibrium, surrounded by a smaller density beam halo. We define the halo as the nonthermal part of the beam that consists of particles having oscillation amplitudes that are larger than the maximum of the beam core.

We observe to have different regimes depending on the value of time scales. There is first a phase of violent relaxation on a time scale $s=50$  leading to a quasi-stationary state. This phase is followed by a thermalization leading to the $T_{eq}$ stationary on a time scale $s=120$ due to the thermal bath (field waves) — i.e., the combined effect of imposed friction and diffusion~\cite{Bohn2,Chavanis4,Tzenov1}, the diffusion process are arising from the fluctuations of the self-fields and the friction results from a polarization process; not  to collisions effects. The first phase is described by the Vlasov-Poisson system with building blocks of the beam core in phase space equipartitioning~\cite{Kandrup3,Chavanis4} and the second phase by the Lenard-Balescu-Poisson system~\cite{Chavanis4,Tzenov1}. But there is a time-scale separation between the phase of violent relaxation and the phase of thermal bath relaxation. We can consider for intermediate times $s=50$ to $s=120$ that the distribution function is a quasistationary solution of the Vlasov equation of the form $f= f(E,t)$ that slowly evolves under the action of imposed friction and diffusion thermal bath, not collisions. Therefore, the system first reaches a state of mechanical equilibrium through violent relaxation , then a state of thermal equilibrium through the effect of imposed fluctuation and dissipation — i.e., the thermal bath. The study this dynamic will be considered in a future work.

\begin{figure}[htb]
\centering
\includegraphics*[width=85mm]{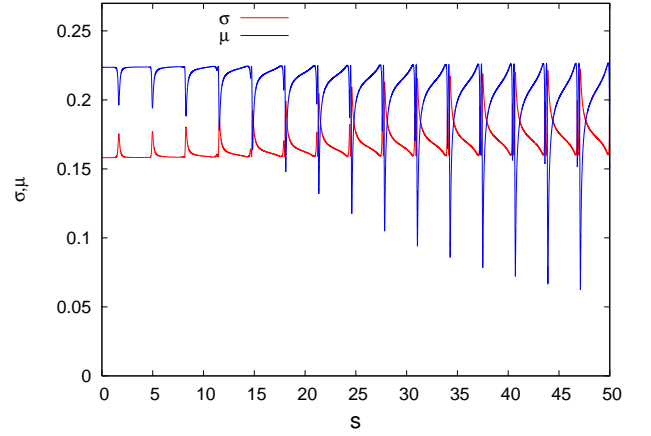} 
\caption{Evolution of the volume and surface area obtained of the equations~(\ref{eq13}) and (\ref{eq14}), respectively. $\mu$ and $\sigma$ are represented by red and blue lines, respectively.}
\label{volume1}
\end{figure}

The dynamics of particles described by these processes has a complex phase space structure. The effectively accessible phase space of the beam can have a complicated geometrical structure. Qualitatively, the accessible phase space of the beam can be computed by volume and surface area that the beam occupies in phase space. We can then measure the volume and surface area of phase space spanned by beam. A commonly employed measure of volume is the root-mean-square geometric emittance $\mu=\sqrt{\epsilon_x\epsilon_y}$ and of the surface area is the sum of the emittance $\sigma=\epsilon_x + \epsilon_y$~\cite{Bohn3}. To the parabolic beam of the Sec.~\ref{EQM} the volume $\mu$ and surface area $\sigma$ are calculated analytically using the emittance equation~(\ref{eq5}). The surface area $\sigma$ is given by :
\begin{equation}
\label{eq13}
\sigma=\frac{1}{90}\sqrt{15}KA\left[\frac{(\chi^2+1)\left(5\chi^2+2\chi+5\right) }{\left(1+\chi\right)^4}\right]^{1/2}
\end{equation}
and the volume $\mu$ is given by :
\begin{equation}
\label{eq14}
\mu=\frac{1}{90}\sqrt{15}KA\left[\frac{\left(5\chi^3+2\chi^2+5\chi\right) }{\left(1+\chi\right)^4}\right]^{1/2}
\end{equation}
where $K$ is the dimensionless perveance of the beam, $A$ is mismatched amplitude of the oscillatories modes beam and $\chi=r_x/r_y$ is ratio of the envelope beam. The temporal evolution of the volume $\mu$ and surface area $\sigma$ of the beam are shown in Fig.~\ref{volume1}.

As shown in Fig.~\ref{volume1} the dynamics of $\mu$ and $\sigma$ present jumps. They are coupled because when the evolution of the volume $\mu$ is maximum, the evolution of the surface area $\eta$ is a minimum and vice versa, -i.e. the particles oscillate between the center and surface of the beam in real space $X-Y$ and in the phase space, when the evolution of the surface area is minimal, the particles are distributed along the volume and when the evolution of the surface area is maximum, the particles are distributed on the surface gaining velocity.

\begin{figure}[htb]
\centering
\includegraphics*[width=85mm]{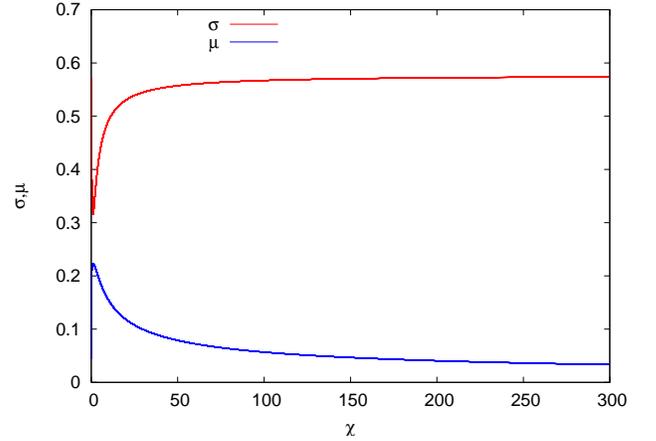} 
\caption{Volume and surface area as a function of anisotropy variable $\chi=r_x/r_y$. $\sigma$ and $\mu$, versus $\chi$ are represented by red and blue lines, respectively.}
\label{volume2}
\end{figure}

In the Fig.~(\ref{volume2}) the volume $\mu$ and surface area $\sigma$ vary in function of $\chi$. For $\chi <1$ $\mu$ increases and $\sigma$ decreases. At $\chi = 1$ the surface area has the minimum value and the volume has maximum value. For $\chi> 1$ $\mu$ decreases and $\sigma$ increases. At $\chi \approx 150$ surface area and volume become stationaries . Note that the surface area is always greater than the volume in  function of anisotropy variable $\chi$.

The volume $\mu$ and surface area $\sigma$ are not constant in function of the axial coordinate of the beam $s$. But the ratio of oscillations energies in the $x$ and $y$ directions $\xi$ remaines constant in function of the $s$ as illustrated in the Fig.~\ref{aniso}. The $\xi$-space is the relevant one to make the statistical mechanics of anisotropic beam. Arguably, the only crucial point is the possibility of constructing time-independent solutions to the Vlavov equation which depend on quantities other than global isolating integrals such as energy, so as to ensure that, if initial data be evolved into the future along the characteriztics associated with the self-consistent potential, the form of initial distribution function $f_0$ will remain unchanged~\cite{Yankov1,Kandrup3}.The key point is that, in principle, a self-consistent equilibrium can be constructed from any set of time-independent phase-space building blocks which, when combined, generate the density distribution associated with an assumed time-independent potential.

As the surface area and volume become stationaries in function of the anisotropy variable $\chi$ we can develop an approach to modelling locally anisotropic kinetic processes~\cite{Vacaru1}. The generalization of kinetic equations to anisotropics bams is not a trivial task. In order to consider the anisotropic processes of a microscopic or macroscopic nature it is necessary to reformulate
the kinetic theory in anholonomic frames. The modelling of kinetic processes with respect to anholonomic frames is very useful with the aim of elucidating flows of fluids of particles not being in local equilibrium. The deduction of the equation for one-particle distribution function in anisotropics phase-space will be considered in a future work.

\section{\label{Conclu}Conclusions}

We examined the transition from isotropic to anisotropic beam profiles in a linear focusing channel. We considered a high-intensity ion beam in space-charge dominated regime and large mismatched beam size-rms initial and we observed a fast anisotropy situation of the beam characterized for a transition by the transversal section from round to elliptical with the coupling of transversal emittance.

The coupled motion between the two tranverse coordinates of a particles beam is arising from the space-charge forces. A beam with nonuniform charge distribution always gives rise to coupled motion. However, it is only when the ratio $r_x/r_y = 1$ and large beam size-rms mismatched that the coupling produces an observable effect in the beam as a whole. This effect arises from both a beating in amplitude between the two coordinate directions for the single-particle motion and from the core-core resonances, resulting in growth and transfer of emittance from one phase plane to another, in the beam developing an elliptical shape along directions preferential and therefore in the fast transition from isotropic to anisotropic beam profiles.

Nonlinear space-charge forces lead to equipartitioning of energy between degrees of freedom. In space-charge dominated beams, Coulomb collisions are infrequent to account for the energy transfer, whereas space charge waves have been shown to be a possible coupling canditate~\cite{Wangler2,Kishek1,Kishek2}. The equipartitioning of anisotropic beam involves nonlinear energy transfer and evolution towards a metaequilibrium state, as consequence of resonant phase mixing~\cite{Bohn1,Kandrup1,Kandrup2}. We used FFTW (The Fastest Fourier Transform in the West) in order to compute frequencies numerically. The breathing mode frequency is given by $\omega_{X_s}=2.01334$ and the quadrupole mode frequency is $\omega_{X_a}=2.01334$. Both modes have the same frequency, therefore breathing mode and quadrupole mode are resonants. This core-core resonance are making the beam to develop an elliptical shape with a increase in its size along the $x$-direction~\cite{Hofmann2}.  We computed numerically the frequencies $\omega_x$ and $\omega_y$ of $2500$ test particles. These $2500$ test particles have $\omega_x=1.05461$ and $2455$ test particles have $\omega_y=1.05461$. Thus $2455$ test particles obey resonance condition $\omega_x-\omega_y=0$. This number large of test particles in resonance causes emittance coupling. As shown in the histogram in Fig.~\ref{fft}, $45$ test particles have $\omega_y=1.00667$ . The particle-core (2:1) resonances are present in both quadrupole and breathing mode of the core oscillation, but only in $y$-direction of the test particles~\cite{Ikegami1}. Thus, halo is formed along $y$-direction. The anisotropy leading to coupling resonance~\cite{Hofmann3,Hofmann4} in the presence of nonlinear space-charge forces was suggested as an approach to the equipartitioning question~\cite{Wangler2,Lagniel3,Kishek1,Kishek2}.

The equipartition of the beam is driven for anisotropic processes. Based on turbulent equipartiton~\cite{Yankov2} we propose the anisotropic equipartition. Anisotropic equipartition corresponds to a phase space density uniform on the surface invariant of the ($\xi=(r_y\epsilon_x)^2/(r_x\epsilon_y)^2$), where $\xi$ is the ratio of oscillations energies in the $x$ and $y$ directions, a version of the ergodic hypothesis where the  $\xi$ invariant play the role of the conserved energy~\cite{Kandrup3}. In the state of anisotropic equipartition, the temperature is stationary, the entropy grows in the cascade form, there is a coupling of transversal emittance, the beam develops an elliptical shape with a increase in its size along one direction and there is halo formation along one direction preferential.

In order to understand the kinetic and dynamical behavior of an anisotropic beams, in particular, to study mechanisms of equipartition connected with phase space, we compute the variables (volume, area of surface, area projected, temperature, heat flux and beam entropy) which characterize the anisotropic beam in phase space.  We observed transverse phase planes coupling in $x-x^{'}$ and $y-y^{'}$ area projected defined by $a_x$ and $a_y$. This coupling is characterized by the energy exchange between the directions. Energy/acceptance exchange requires resonant coupling. Energy/acceptance flow is due to the correlation $jj^{'}$ between the velocity and the position of the particle in regions where the beam size contracts or expands. When the beam radius expands, the flow component has an outward direction and the thermal component decreases, when the beam radius contracts, the flow component is inward and the thermal component increases. The propagation of the beam is thus characterized by the variation in the envelope beam which correlates with the generation of the heat flow and the variation of the beam temperature. The elementary mechanism responsible for the heat transfer from one degree of freedom to another is constituted by the effect of interactions between the individual particles and waves of collective fields~\cite{Elskens1}.

The temperature $T_{eq}$ oscillates after steady with small fluctuations and entropy $S$ grows by cascades as shown in Fig.~\ref{lyapu}. We observe to have different regimes depending on the value of time scales. There is first a phase of violent relaxation leading to a quasi-stationary state. This phase is followed by a thermalization leading to the $T_{eq}$ stationary due to the thermal bath (field waves) — i.e., the combined effect of imposed friction and diffusion~\cite{Bohn2,Chavanis4,Tzenov1}, the diffusion process are arising from the fluctuations of the self-fields and the friction results from a polarization process; not  to collisions effects. The particles dynamic described by these processes has a complex phase space structure. The effectively accessible phase space of the beam have a complicated geometrical structure. We showed that the dynamics of the volume $\mu$ and the surface $\sigma$ are coupled. They are not constant in function of the axial coordinate of a beam $s$. But the surface area and volume become stationaries in function of the anisotropy variable $\chi$.

This series of phenomena suggest an algorithm for future research. First, to map the building blocks of the beam core in the phase space that energy equipartition of the beam. The dynamic is described as a trajectory in the phase space. If the system has some integrals of motion, the trajectory is confined in the subspace which conserves the integrals. The system is defined as ergodic if the time average of any dynamical quantity is equal to the spatial average over the subspace. This subspace is referred to as the ergodic region. At the equilibrium, energy is equally distributed to all degrees of freedom, which is called equipartition. In particular, this anisotropic beam has a quasi-stationary state that seems to suggest the existence of additional integrals, which conserves the anisotropy~\cite{Kandrup3,Zeeuw1}. The anisotropy lead to the equipartition beam.
 
Second, to describe self-consistently, the interactions between the individual particles and waves of collective fields via hamiltonian formalism~\cite{Elskens1}, these interactions form the halo. The traditional description of the interacting particles and fields rests on the coupled set of Vlasov-Poisson equations. In the self-consistent approach to the wave-particle interaction complements this usual treatment in that, to capture the physical mechanism at work, the beam is partitioned in two populations: core and tail. The idea behind this discrimination is simple: wave-particle interaction involves the resonant tail particles whose velocity is close to the phase velocity of the wave under consideration. These waves are just the collective macroscopic degrees of freedom, capturing the oscillations of other nonresonant, core particles, so that these core particles participate in the effective wave-particle dynamics only through the waves.

Third, to analyze the halo-core interaction via thermal bath, ie, the Fokker-Planck equation. The Lenard–Balescu equation can be used to describe the evolution of the halo particle in a bath of field particles (core) at equilibrium. In that case, we have to consider that the distribution of the bath is given, that is $f(v,t)=f_0(v)$ where $f_0(v)$ is a stable stationary solution of the Vlasov equation (bath distribution). This procedure transforms the Lenard-Balescu equation into a Fokker-Planck equation for the density probability $P(v,s)$ of finding the halo particle with velocity $v$ at time $s$. Chavanis~\cite{Chavanis2} studied the relaxation of a test particle in a bath of field particles. Specifically, he considered a collection of $N$ particles at statistical equilibrium (thermal bath) and introduced a new particle in the system. These results are well-known when the system is spatially homogeneous and memory effects can be neglected, as in the case of plasma physics.

\begin{acknowledgments}
Regrettably, my father Wilson Simeoni died on 15 November 2008. This paper stands as a tribute to his myriad contributions to my life. The author would like to thanks Conselho Nacional de Desenvolvimento Cient\'{\i}fico e
Tecnol\'ogico (CNPq), Brazil and PAC07 Local Organizing Committee (LOC), mainly, Tsuyoshi Tajima for the financial support.
\end{acknowledgments}


\newpage 

\end{document}